\def\kp  {$K^{\prime}$}
\def\kms{\ifmmode{\hbox{km~s}^{-1}}\else{km~s$^{-1}$}\fi}
\def\la{\mathrel{\hbox{\rlap{\hbox{\lower4pt\hbox{$\sim$}}}\raise1pt\hbox{$<$}}}}
\def\ga{\mathrel{\hbox{\rlap{\hbox{\lower4pt\hbox{$\sim$}}}\raise1pt\hbox{$>$}}}}
\documentstyle[aaspp4]{article} 
\begin{document}
\title {\bf Distances to Galaxies from the Correlation Between Luminosities
and Linewidths.  III. Cluster Template and Global Measurement of H$_0$.}

\author{R. Brent Tully$^1$ and Michael J. Pierce$^2$}

\affil {$^1$ Institute for Astronomy, University of Hawaii, 
2680 Woodlawn Drive, Honolulu, HI 96822 e-mail: tully@ifa.hawaii.edu}
\affil {$^2$ Department of Astronomy, Indiana University,
Swain West 319, Bloomington, IN 47405 e-mail: mpierce@astro.indiana.edu}

\begin{abstract}

The correlation between the luminosities and rotation velocities of galaxies
can be used to estimate distances to late-type galaxies.  It is an 
appropriate moment to re-evaluate this method given the great deal of new 
information available.  The major improvements described here include: 
(a) the template relations can now be 
defined by large, complete samples, (b) the samples are drawn from a wide
range of environments, (c) the relations are defined by photometric information
at $B,R,I$ and \kp\ bands, (d) the multi-band information clarifies 
problems associated with internal reddening, (e) the template zero-points
are defined by 24 galaxies with accurately known distances, and (f) the
relations are applied to 12 clusters scattered across the sky and out to
velocities of 8,000~\kms.  The biggest change from earlier
calibrations are associated with point~(e).  Roughly a 15\% increase
in the distance scale has come about with the five-fold increase in the
number of zero-point calibrators.  The overall increase in the distance 
scale from the luminosity--linewidth methodology is about 10\% after
consideration of all factors.  
Modulo an assumed distance to the Large Magellanic Cloud of 50~kpc and
no metallicity corrections to the Cepheid calibration, the resulting value 
of the Hubble Constant
is H$_0 = 77 \pm 8$~\kms~Mpc$^{-1}$ where the error is the 95\% probable
statistical error.  Cummulative systematic errors internal to this
analysis should not exceed 10\%.  Uncertainties in the distance scale
ladder external to this analysis are estimated at $\sim10\%$.
If the Cepheid calibration is shifted from the LMC to NGC~4258 with a
distance established 
by observations of circum-nuclear masers then H$_0$ is larger by 12\%.

\end{abstract}


\section{The Situation is Improved}

There have been substantial advances with both the quality and quantity
of methods used to determine the distances to galaxies.  The procedures
now available have an interesting mix of complementary strengths.
{\it Cepheid variable stars} in galaxies with young populations can be
observed effectively with Hubble Space Telescope (HST) out to distances
of 30~Mpc, though only modest numbers of galaxies can afford to be 
targeted (Kennicutt, Freedman, \& Mould 1995). \markcite{ke1}  The 
characteristic 
luminosity of the {\it tip of the red giant branch} provides an accurate 
estimator of distances to old, metal poor populations.  Red giant stars 
are resolved with HST for galaxies within 15~Mpc (Madore \& Freedman
1995, \markcite{ma1} Harris et al. 1998. \markcite{ha}  Even if these 
stars are unresolved or 
blended by crowding, their statistical properties provide distance
estimates through characterization of {\it surface brightness
fluctuations}.  Distances to early-type galaxies are 
effectively measured to 40~Mpc at $I$-band (Tonry et al. 1997,1999)
\markcite{to} \markcite{to2} and the 
method is being pushed to twice as far with infrared observations
(Jensen, Tonry \& Luppino, 1998). \markcite{jen}  The {\it planetary nebula
luminosity function} can be established in early-type galaxies within 
30~Mpc (Jacoby, Ciardullo, \& Ford 1990). \markcite{jac}  
{\it Type~I$a$ supernovae} 
are demonstrated to provide excellent distances to a mix of galaxy types
(Riess, Press, \& Kirshner 1996). \markcite{ri}  The supernova method can be 
applied to great distances but local coverage is sparse.  All of
the above methods evidently can provide relative distances good to
5-10\% rms.  Absolute scales have zero-point uncertainties of 
comparable amounts.

These techniques provide accurate individual distances but either the
reach is limited, or application is restricted to early types, or
coverage is serendipitous and sparse.  In contrast, the  correlation 
between the luminosity and rotation linewidth of spiral galaxies (Tully \&
Fisher 1977) \markcite{tu3} provides a distance measure that is not
as accurate per object, 15-20\% rms, but {\it it can be applied to 
thousands of 
galaxies} out to 100~Mpc and beyond.  Roughly 40\% of galaxies with
$M_B<-16^m$ have appropriate morphologies and orientations, so are 
potential targets.

The luminosity--rotation linewidth method has been used often
($cf,$ Aaronson et al.  1979, \markcite{aa1} 1986; \markcite{aa2} 
Bottinelli et al.  1983; \markcite{bo1} Pierce \&
Tully 1988, \markcite{pt1} 1992; \markcite{pt2} Mould et al.  1993; 
\markcite{mo} Giovanelli et al.  1997$b$; \markcite{gi3} Willick et al. 
1996; \markcite{wi2} Willick \& Strauss 1998). \markcite{wi4} 
Recent progress has lead to four major 
improvements. In Paper I of this series (Pierce \& Tully 1999$a$), 
\markcite{pt3} photometry at $B,R,I$ bands are presented 
for galaxies that have distance estimates based on the Cepheid 
period--luminosity relationship and which provide the absolute 
calibration of the luminosity--rotation linewidth relations.  In Paper II 
of this series (Pierce \& Tully 1999$b$), \markcite{pt4} $B,R,I$ 
photometry are presented for three substantial cluster samples in 
order to examine the form of the relations and determine distances 
to these clusters.  In the present paper, this material is 
combined with extensive $I$-band material from the literature and
more limited \kp -band data.  The large number of $I$-band observations
leads to two of the four significant improvements.  {\it Improvement one: 
the template
luminosity--linewidth correlation is now defined by samples that are
statistically well defined, substantial, and drawn from a wide range
of environments.}  The template relation is used to obtain relative
distances to twelve clusters.  {\it Improvement two: the Hubble
parameter is measured at a statistically significant number of locations
around the sky in the redshift range 3,000-8,000~\kms.}  The 
multi-band information provides partially independent distance
estimates.  The \kp\ measures are essentially free of reddening 
uncertainties.  {\it Improvement three: corrections for inclination 
effects are better understood.}  Finally, and the source of the 
biggest change, the HST observations of Cepheid variables has enabled 
accurate distance measurements to many more galaxies suitable for 
calibrating the luminosity--linewidth correlation.  {\it Improvement 
four: there are now a substantial number of zero-point calibrators.}

\section{Data}

The application of the luminosity--linewidth correlation requires the 
measurement of three parameters: an apparent magnitude, a characterization
of the rotation rate, and an estimate of the inclination needed to 
compensate for projection effects.  The measurement of these components 
will be considered in turn.  Then, there will be a discussion of the 
adjustments to be made to obtain the parameters that are used in the 
correlations.

\subsection{Luminosities}

Large format detectors on modest sized telescopes provide fields of 
view that can encompass essentially any nearby galaxy.  As a result, surface 
photometry 
with optical and near-infrared imagers is now relatively routine.  However, 
the low surface brightness of galaxies compared with the night sky still 
presents significant challenges for accurate photometry.  The authors have 
an on-going program of both optical ($B,R,I$) and infrared (\kp) photometry 
(Pierce \& Tully 1988, 1992, 1999$a,b$; Tully et al. 1996, \markcite{tu8} 1998; 
\markcite{tu5} Rothberg et al. 1999: hereafter RSTW). \markcite{rst} Our 
observations of 
relevance to the current analysis pertain to the three clusters Ursa Major,
Pisces, and Coma (see Paper II) and to nearby 
`calibrators' with independently established distances (see Paper I).  
Three-band optical data are available for 
all four of these 
separate samples and \kp\ data are available for UMa, Pisces, and a limited
number of calibrators.

There is also now a wealth of $I$-band photometry material in the literature.
There are good overlaps for comparisons between the major sources, especially
if we do not 
restrict ourselves to just the samples required for the present analysis.
For the purposes of this paper, the
important sources of luminosities other than our own are Mathewson,
Ford, \& Buchhorn (1992), \markcite{mat} Han (1992), \markcite{han} 
Bernstein et al. (1994), \markcite{ber}
Bureau, Mould, \& Stavely-Smith (1996), \markcite{bu} and Giovanelli et al. 
(1997$a$). \markcite{gi4}
These five sources provide $I$-band  magnitudes for galaxies in clusters
at intermediate to large distances (to $\sim 8,000$~\kms).   The inclusion
of this material permits the construction of an $I$-band luminosity--linewidth
template based on five clusters (Fornax and Abell~1367 in addition to the 
clusters mentioned above).  The template calibration can then be fit to
give distances to seven more clusters.

At the moment, there are a lot more data available at $I$ than at other bands
so most of the analysis presented in this paper will be based on this material.
There is interest in the other bands, though, because of the insidious effects
of obscuration.  It should provide comfort that there is proper compensation
for these effects when it is demonstrated that relative distances are the same 
at different passbands.
The $K^{\prime}$ material is of particular interest in this regard since
obscuration should be very small at 2.1~microns.

The issue of adjustments to magnitudes because of obscuration and spectral
shifting will be discussed in a later section.  The concern at this point is
the homogeneity of the raw magnitudes from various sources.  Different
authors measure magnitudes to slightly different isophotal levels then usually
extrapolate to total magnitudes:
Han (1992) extrapolates from $I_{iso}=23.5^m$, Giovanelli et al. (1997$a$) 
extrapolate from $\sim 24^m$, Mathewson et al. (1992) extrapolate from 
$25.0^m$, Tully et al. (1996) extrapolate
from $25.5^m$, and Pierce \& Tully (1999$a,b$) extrapolate from $26.0^m$.
The added light at the faintest levels is small for the 
high surface brightness galaxies that are relevant for the determination of
H$_0$.  For luminous, high surface brightness galaxies typical extrapolations 
from $I_{iso}=25.5^m$ to infinity add $\sim0.02^m$ and for the faintest, low 
surface brightness systems the corrections are still less than $0.1^m$ 
(Tully et al. 1996, Papers I, II).  Magnitude measurements are sensitive 
to the depth of the surface photometry and the detailed fitting of the sky 
level and variations at the level of $\sim 0.05^m$ are common.

Inter-comparisons between sources indicate that the various sources cited 
here are all on the same system and that systematic errors are almost 
negligible.  Some offsets between data sets have been reported, for example 
Giovanelli et al (1997$a$) adjust Mathewson et al. data (1992) to match 
their own.  However, the data sets are consistent with each other at a 
level of 4\%, or 2\% in derived distances.  The object by object,
rms differences between any pair of observers is at or below $\pm 0.1^m$.
In the present analysis all sources are given
equal weight and luminosities are averaged if there are multiple
observations.  Overlap measurements do reveal spurious results in a few
percent of the cases.  If a difference between sources is large it
is usually evident which measurement is incorrect.

\subsection{Inclinations}
Projection corrections are required to recover true disk rotation rates
and to compensate for differential obscuration.  Uncertainties in inclination
increasingly affect de-projected velocities as one approaches face-on 
orientation.  With rare exception, inclinations are derived from a
characteristic axial ratio of the main or outer body of a galaxy.  From
experience, it is found that such inclination measurements are reproducable
at the level of $\pm 3^{\circ}$ rms.  However, the errors are non-gaussian.
From the radial variations in axial ratios and from such independent 
considerations as inclination estimates derived from two-dimensional velocity 
fields it is suspected that errors as large as $\sim 10^{\circ}$ are not 
uncommon.  In this case, the $1/{\rm sin}i$ deprojection correction becomes 
very uncertain toward face-on orientation.   We apply a sample cut-off at 
$i=45^{\circ}$ to avoid large errors.

The derivation of an inclination from an axial ratio requires an assumption
about the intrinsic thickness of the system.  The standard formulation
(Holmberg 1958) \markcite{ho} is ${\rm cos}i = \sqrt{(q^2-q_0^2)/(1-q_0^2)}$ 
where $q=b/a$ is the observed ratio of the minor to major axes and $q_0$ is 
the intrinsic axial ratio.  The thinnest systems are spirals of type Sc.
Earlier types have bulge components and the disks of later types are less 
flattened.  For simplicity, a single 
value for the the flattening is often used and $q_0=0.20$ will be used 
in this analysis.  A more elaborate specification of $q_0$ that depends on 
the morphological type could be justified.  Giovanelli et al. (1997$a$) 
provide an extreme example with their choice $q_0=0.13$ for type Sc.  All 
other measurements being equal, a smaller $q_0$ value results in derived 
inclinations that are more face-on.  Fortunately
the choice of $q_0$ has a negligible effect on the measurement of distances 
as long as one is consistent between the calibration and subsequent target 
samples.  
For an observed $q=0.20$, the difference $q_0=0.13$ or 0.20
gives a difference in inclination of
$81^{\circ}$ or $90^{\circ}$ respectively.  However the $1/{\rm sin}i$ 
difference on the corrected linewidth is only 1.2\%.  As one progresses
toward larger $q$ the difference in assigned inclination is reduced but the
$1/{\rm sin}i$ correction is growing.  The product of the two effects is a
roughly constant shift of 1.2\% in the corrected linewidth at all inclinations 
$i>45^{\circ}$.  If both calibrators and subsequent targets are handled in the 
same manner there will be no significant effect on the measured distances.

Extinction corrections due to projection affect luminosities 
in the opposite inclination regime.  The correction is highest for edge-on 
systems and 
decreases as galaxies are presented more edge-on.  It has become traditional 
to formulate extinction corrections directly in terms of the observed $q$ 
value which avoids a dependence on the parameter $q_0$.  This approach is
adopted here, as discussed in section 2.4.

Inter-comparisons between the various sources of photometry used in this study
do not reveal any systematic differences in $q$ measurements between authors.
Large individual differences are not uncommon, often associated with
systems with pronounced non-axisymmetric structural features such 
as bars.  Deep surface photometry can usually allow the origin of these 
discrepancies to be identified and allow the appropriate choice of $q_0$ 
to be made.
 
\subsection{Linewidths}
It is possible to measure rotation parameters via both optical and radio
techniques.  The original radio methods are simpler but are constrained by
detector sensitivity to modest redshifts.  The methods that involve optical
spectra require more work but can be used to larger distances.  With care, 
the two techniques can be reconciled into a common characterization of the
projected rotation speed (e.g. Courteau 1997). \markcite{co}  However that 
synthesis will not be attempted here.  Observations of Doppler-broadened 
profiles in the 21~cm neutral hydrogen line are available for galaxies at 
distances adequate for the purpose of determining H$_0$.  We will avoid 
the added complexity of intermingling radio and optical data.

Even if linewidth measurements are limited to HI data, there are still 
complications.  For once in astronomy, angular resolution is not an 
unmitigated advantage.  Specifically, it is necessary that the velocity 
field be sampled well out onto the flat portion of the rotation curve.  
Modern synthesis techniques can over-resolve the velocity field and 
result in decreased sensitivity to fainter, extended emission.  As a 
result, the data on nearby, large galaxies handed down from observations 
on old telescopes from the days of paper strip charts is still preferred.  
However, the measurements then remain somewhat `personalized'.  Even more 
than with magnitudes, one has to be careful to use a consistent set of 
linewidth information from the near field to the far field.  In this study, 
HI profile linewidths defined at the level of 20\% of the peak flux are 
used (called $W_{20}$).  These linewidths are only adequately measured 
if the signal-to-noise ($S/N$) at the emission line peak is greater than 7.  
A clean profile with $S/N>10$ typically provides a measurement of $W_{20}$ 
with an accuracy of better than 10~km s$^{-1}$.  Mediocre profiles 
($7<S/N<10$) have linewidth uncertainties of $\sim 15$~\kms.  Profile 
measures with uncertainties $>20$~\kms\ are not accepted.  The 20\% 
linewidths are then corrected for projection and internal turbulence 
resulting in the parameter $W_R$ defined by Tully \& Fouqu\'e (1985). 
\markcite{tu4} This parameter is constructed to approximate twice the 
maximum rotation velocity of a disk galaxy.

An alternative linewidth characterization in common use is the width at 
50\% of peak flux in each horn of the profile ($W_{50}$: Haynes et al. 1997) 
\markcite{ha3} which is then adjusted to account for instrumental and 
thermal broadening (Giovanelli et al.  1997$a$).  The advantages and 
disadvantages of the alternative systems are technical and not very 
important.  The key concern is that the information from both 
northern and southern hemispheres and for both nearby large
galaxies and those distant and small be brought to a common system.
The current analysis draws on a database of $W_{20}$ measurements for 4500
galaxies within 3000~\kms\ maintained by the first author.  Linewidths
for the more distant clusters have been obtained from the literature or
from M.P. Haynes (private communication) and measured in a consistent way.

\subsection{Extinction Corrections}
A significant improvement in the present calibration results from new
corrections for internal extinction.  Giovanelli et al. (1995) 
\markcite{gi5} made a convincing case for a strong luminosity
dependence in the internal obscuration of galaxies and Tully et al. (1998)
have further quantified the effect.  The latter work has profited from
the leverage provided by information in passbands from $B$ to $K^{\prime}$.
At edge-on orientation a giant galaxy can be dimmed by 75\% at $B$, while 
the extinction within a dwarf galaxy with the luminosity of the Small 
Magellanic Cloud cannot be statistically measured.  For comparison, at 
$K^{\prime}$ the most luminous galaxies are dimmed by a maximum of only 20\%.

The inclination-dependent extinction can be described by the expression 
$A_i^{\lambda}=\gamma_{\lambda}{\rm log}(a/b)$ where $a/b$ is the major to
minor axis ratio and $\lambda$ is the passband.  The correction is to
face-on orientation and hence does not account for the residual absorption 
within a
face-on system.  Given the strong luminosity dependence, there is a potential 
problem since the absolute magnitudes are not known a priori.  Absolute 
magnitudes are to be an output of the distance estimation process so they
cannot also be an input.  Both Giovanelli et al. (1997$a$) and Tully et al. 
(1998) recast the corrections for magnitudes so the dependency is on the
distance-independent linewidth parameter.  This conversion is provided
through the luminosity--linewidth calibrators.  The formulations presented by
Tully et al. (1998) are:
\begin{equation}
\gamma_B = 1.57 + 2.75 ({\rm log} W_R^i - 2.5)
\end{equation}
\begin{equation}
\gamma_R = 1.15 + 1.88 ({\rm log} W_R^i - 2.5)
\end{equation}
\begin{equation}
\gamma_I = 0.92 + 1.63 ({\rm log} W_R^i - 2.5)
\end{equation}
\begin{equation}
\gamma_{K^{\prime}} = 0.22 + 0.40 ({\rm log} W_R^i - 2.5)
\end{equation}
There is a fortunate interplay that minimizes the effect of uncertain
inclination on $A_{\lambda}$.  If the inclination is taken too face-on because 
of a spuriously
large $b/a$ then $W_R^i$ is overestimated, which drives up $\gamma_{\lambda}$,
but is offset by a low ${\rm log}(a/b)$ in the product that gives 
$A_{\lambda}$.  Here, $W_R^i \sim 2 V_{max}$ where $V_{max}$ is the amplitude of
maximum rotation in a galaxy (Tully \& Fouqu\'e 1985).

The luminosity dependencies found by Giovanelli et al. (1995) and Tully at el.
(1998) are similar so it had been expected that the inclination corrections 
advocated here would be similar in amplitude to the corrections used in
recent papers by Giovanelli and collaborators.  However, in fact, there
is poor agreement.  The correction reformulation in terms of linewidths
offered by Giovanelli et al. (1997$a$) has a consequence that seems unintended
by those authors.  The average correction is largest at 
$2 V_{max} \sim 225$~\kms\ 
and on average then {\it decreases} progressively toward higher linewidths.
By contrast, our corrections increase continuously toward higher linewidths.  In 
order to evaluate the effect of alternative extinction corrections on the determination 
of H$_0$, we have carried three formulations through all stages of the 
calibration: the corrections we advocate, the corrections described by 
Giovanelli et al. (1997$a$), and corrections with no linewidth or 
luminosity dependency.  The slopes of the luminosity--linewidth correlations
are very different according to the choice of correction algorithm and
scatter is dependent on the choice (lowest with our corrections) but 
as long as consistency is maintained throughout the analysis the final
overall distance scales are the same in all three cases to within 0.5\%.  
Evidently, reasonable changes in the extinction correction procedure have 
negligible effect on H$_0$.

The other corrections to be made are modest and non-controversial.
Galactic absorption was calculated from the 100 $\mu$m cirrus maps of 
Schlegel, Finkbeiner, \& Davis (1998) \markcite{scl} according to the
reddening curve description 
$A_b^{\lambda}=R_{\lambda}E(B-V)$, $R_{\lambda}=4.32$, 2.68, 1.77, 0.37
for $\lambda=B,R,I$,\kp.
We make a small $k$-correction
at both $I$ and $R$ bands of $A_k^{R,I}=(4.24(R-I)-1.10)z$.  At $B$ we
make the correction $A_k^B=(3.6-0.36T)z$ where T is the galaxy morphological
type in the familiar convention: T:1,3,5,7=Sa,Sb,Sc,Sd.  These 
$k$-corrections are always $<0.08^m$ at $B$, $<0.04^m$ at $R,I$, and
negligible at \kp\ for the current samples.

\subsection{Data Summary}
The averaged data are accumulated in Table~1.  The following information is provided
in each column. (1) {\it Top:} 
Names; by preference, NGC (N), UGC (U), IC, Zwicky (Z), ESO-Uppsala (E),
or, in the 9 cases without PGC designations, the identification number is
from table 2 of Giovanelli et al. (1997$a$). 
(1) {\it Bottom:} Principal Galaxies Catalogue (PGC) number from the 
Lyon Extragalactic Database (available for all but 9 cases).
(2-4) For the 24 zero-point
calibrators, the accepted distance moduli are given here.  For the cluster
galaxies, the successive columns provide 
equatorial coordinates
(epoch 1950), galactic coordinates, and supergalactic coordinates.
(5) Morphological types (T:1,3,5,7,9=Sa,Sb,Sc,Sd,Sm). 
(6) {\it Top:} Systemic
velocity in the rest frame of the cosmic microwave background.
(6) {\it Bottom:} Axial ratio of minor axis to major axis, $q$. 
(7) {\it Top:} Galactic foreground reddening, $E(B-V)$.
(7) {\it Bottom:} Inclination, $i$.
(8-11) {\it Top:} Total magnitudes, $B_T$, $R_T$, $I_T$, \kp$_T$.
(8-11) {\it Bottom:} Total magnitudes adjusted for galactic extinction ($b$), 
inclination-dependent extinction ($i$), and $k$-correction ($k$), 
$B_T^{b,i,k}$,$R_T^{b,i,k}$,$I_T^{b,i,k}$, and \kp$_T^{b,i,k}$.
(12-15) {\it Top:} Absolute magnitudes at the indicated distance modulus,
$M_B^{b,i,k}$, $M_R^{b,i,k}$, $M_I^{b,i,k}$, and $M_{K^{\prime}}^{b,i,k}$.
(13) {\it Bottom:} HI linewidth, $W_{20}$. 
(14) {\it Bottom:} Linewidth uncertainty.
(15) {\it Bottom:} Logarithm of adjusted linewidth, ${\rm log} W_R^i$.
(16) {\it Top:} References for $I$-band photometry and, in the case of the
calibrators, of distances.
(16) {\it Bottom:} References for HI linewidths.

All $B$ and $R$ magnitudes are from Papers~I and II (Pierce \& Tully 1999$a,b$).
The \kp\ data for the Ursa Major Cluster were presented and discussed by
Tully et al. (1996) and the Pisces and calibrator \kp\ material are drawn
from RSTW.  The $I$ magnitudes are averaged over data from Papers~I and II 
and the other sources identified in column 16 with the following codes:
1 $=$ Papers~I and II, 2 $=$ Tully et al. (1996), 3 $=$ Giovanelli et al. (1997$a$),
4 $=$ Han (1992) and Han \& Mould (1992), \markcite{hm} 5 $=$ Mathewson et 
al. (1992), 
6 $=$ Bernstein et al. (1994), and 7 $=$ Bureau et al. (1996).

HI linewidth references are given by a 3 figure code.  If the code is less
than 600 then the reference is provided by Huchtmeier \&  Richter (1989)
\markcite{hu}
for that code.  We have been maintaining a database that follows on from
Huchtmeier \&  Richter and the additional references of concern are given
here: 601 $=$ Begeman (1989), \markcite{beg}
604 $=$ Gavazzi (1989), \markcite{gav} 613 $=$ Cayatte et al. (1990), 
\markcite{ca}
615 $=$ Fouqu\'e et al. (1990), \markcite{fo} 619 $=$ Magri (1990),
\markcite{mag}
620 $=$ Puche et al. (1990), \markcite{pu} 623 $=$ Schneider et al. (1990), 
\markcite{sn1}
630 $=$ Haynes \& Giovanelli (1991$a$), \markcite{ha1} 631 $=$ Haynes \& 
Giovanelli (1991$b$), \markcite{ha2} 637 $=$ Roth et al. (1991),
\markcite{rot} 
653 $=$ Rood \& Williams (1992), \markcite{roo} 655 $=$ Schneider et al. (1992),
\markcite{sn2}
658 $=$ Mathewson et al. (1992), 660 $=$ Broeils (1992), \markcite{bro}
672 $=$ Bosma \& Freeman (1993), \markcite{bos} 673 $=$ Braine et al. (1993),
\markcite{bra}
686 $=$ Scodeggio \& Gavazzi (1993), \markcite{sco} 691 $=$ Garcia-Barreto 
et al. (1994), \markcite{gar}
699 $=$ Bureau et al. (1996), 
700 $=$ Wegner et al. (1993), \markcite{we}
701 $=$ Haynes et al. (1997), 702 $=$ Haynes, private communication,
703 $=$ Williams, private communication, 704 $=$ Eder, private communication,
705 $=$ Freudling, private communication, 706 $=$ Giovanelli et al. (1997$a$),
707 $=$ Tully \& Verheijen (1997). \markcite{tu7}

\section{Biases}
Over the years many people have used luminosity--linewidth relations to 
measure distances and there has been controversy.  An extreme view 
has been presented by Sandage (1994$b$).  According to him,
there can be large biases that distort distance measurements and limit the
usefulness of the procedure.  In this section there will be a description of
a way of conducting the analysis that results in unbiased distance estimates
and, hopefully, accurate results.  The method to be described is {\it not} 
the method used by Sandage.

Malmquist (1920) \markcite{mal} discussed a bias that might create a problem with
measurements of distances to objects selected by apparent magnitude. Teerikorpi
(1984) \markcite{te} and Willick (1994) \markcite{wi1} have discussed the problem 
in the present context.
Schechter (1980) \markcite{sce} and Tully (1988$a$) \markcite{tu1} have described 
a procedure that is
expected to {\it nullify} the bias.  That procedure will be summarized
after preliminary remarks about methods that do suffer bias.

An example of when the bias arises is provided by considering the description 
of the luminosity--linewidth correlation given by the regression with errors
taken in magnitudes -- sometimes called the `direct' relation.  Use the
`direct' relation to determine distances to objects. Suppose one considers a 
group.
By the construction of the regression, the brightest galaxies will tend to
lie above the correlation line.  Treated one by one, the
brightest galaxies, drawn from above the mean correlation but assigned the
absolute magnitude of the mean correlation, will be given a {\it closer}
distance than is correct.  As fainter galaxies in the group are sampled,
they progressively sample the true distribution around the mean
correlation, so that the mean distances of the fainter galaxies are larger.
Kraan-Korteweg, Cameron, \& Tammann (1988) \markcite{kr} have shown that 
the measured mean
distance of a group increases as fainter objects are included.  For the same
reason, as one probes in the field to larger redshifts one samples 
progressively only the brightest galaxies, those that tend to be drawn from 
above the mean correlation (Bottinelli et al. 1986). \markcite{bo2} Hence 
one progressively 
assigns erroneously low distances.  Low distances give a high H$_0$.

In an analysis made this way it is imperative that a
correction be made for the bias.  However, to make the correction it is
necessary to have detailed information on the form of the luminosity--linewidth
correlation and the nature of the scatter.  With adequate information, it
is possible to correct {\it statistically} for the bias, though the
trend of deviations with magnitude would persist in the individual 
measurements.  However, when is their adequate information?

Historically, the local velocity anomaly (Tully 1988$a,b$) \markcite{tu2} has caused 
confusion.  If we are correct, most galaxies within 1000~\kms\ in the northern
galactic hemisphere have negative peculiar velocities; ie, the ratio of their
observed velocities divided by their distances give low values for the
Hubble parameter.  Non-linear dynamical models of local structure (Shaya,
Peebles, \& Tully 1995) \markcite{shy} anticipate these low values as a consequence of the
gravity of local structures, but that is another story.  The point to be made
here is that an apparent increase in the Hubble parameter locally {\it might}
be caused by Malmquist bias, as Bottinelli et al. (1986) and Sandage 
(1994$a,b$) \markcite{san1} \markcite{san2} argue, or it {\it might} be a real,
physical effect.  If one
assumes the abrupt increase in the mean Hubble parameter at 1000~\kms\ is 
due to bias then one is driven to justify a huge bias correction and conclude
that H$_0$ has a low value.  It appears, though the details are slim,  that 
Theureau et al. (1997) \markcite{th} follow Bottinelli et al. and Sandage on this path.

Variations on the procedures that require bias corrections are pervasive
(eg, Willick et al. 1997). \markcite{wi3}  For example, a maximum likelihood description
of the relationship (Giovanelli et al. 1997$b$; Watanabe, Ichikawa, \&
Okamura 1998) \markcite{wa} still retains the bias and 
requires corrections.  The corrections might be done properly.
However, these procedures require (1) that the
calibrators and targets have the same statistical properties, and (2)
detailed specification of the sources of scatter
and of properties of the luminosity function from which the sample is
drawn.
As an alternative, the method to be described {\it nulls} the bias 
rather than {\it corrects} for it.
Consequently, there is no
requirement to specify the sources of scatter or the properties of the sample. 
One is relying only on the
assumption that calibrators and targets have the same properties.

The magic description that nulls the bias is given by the regression with
errors in linewidth (Schechter 1980; Tully 1988$a,b$) -- 
the `inverse' relation.  Two
qualitative comments might crystallize the merits of the 
procedure.  The first point to appreciate is that {\it the amplitude of the
bias depends on the assumed slope of the correlation.}  The flatter the 
description of the
dependence of magnitude with linewidth the greater the bias.  Conversely, 
if the slope is taken steep enough {\it the sign of the bias can be reversed.}
Hence it can be understood that there is a
slope that nulls the bias.  That slope is given by the regression on
linewidth if the sample is only limited in magnitude.
The second key point is made by a 
consideration of the regressions on the separate axes of a 
luminosity--linewidth plot.  Suppose one considers successively brighter
magnitude cuts on an intrinsic distribution.  As one progressively limits 
the magnitude range, the correlation coefficient of the fit will degrade.
Presented graphically, the correlations on the two axes will progressively
diverge as the fitting range is reduced.  Here is the critical point.
As the truncation is progressively advanced in magnitude the slope
with errors in magnitudes is progressively splayed to shallower values but
{\it the slope 
of the correlation with errors in linewidths is always the same.}

Since the amplitude of the bias depends on the slope of the correlation, it
should be seen that an analysis based on the direct relation is on slippery
ground because the value of the slope depends on the magnitude limit of
the sample.  One needs a lot of information for an internally consistent
application.  The maximum likelihood approach raises the same qualitative 
concerns although, because it involves a slope intermediate between the
direct and inverse correlations, the quantitative problem is also
intermediate.

It has been pointed out by Willick (1994) that a bias can enter the inverse
correlation in practical applications.  The bias can be introduced because
the cutoff may not be strictly in magnitude.  For example, the sample might
be chosen at $B$ band but applied at a more redward band such as $I$.
A correlation between color and linewidth generates a slope to the magnitude
cutoff at a band other than $B$.  Or suppose the sample is selected by 
apparent diameter.  A correlation between surface brightness and linewidth
can again give a slope to the magnitude cutoff.  A slope in the magnitude
cutoff is equivalent to the introduction of a linewidth stricture.   Any 
restriction in linewidths brings the problem of bias over to the orthogonal
axis.  Two things can be said of this problem.  First it is a small effect,
down compared with the `direct' relation by a factor of five in amplitude 
in Willick's analysis.  Second
the problem is partially avoided by building the calibration out of only 
galaxies that satisfy a completion limit at the band to be considered; ie,
a stricter limit is taken than the one that provided the initial sample.

Most important: to achieve the correlation that nulls the bias one wants 
{\it a complete magnitude limited calibration
sample.}  In the population of the luminosity--linewidth diagram with the
calibration sample there should
not be any discrimination against candidates in any particular part of the
diagram above the magnitude limit.  Selection based on 
inclination is inevitable but that restriction should be distributed across 
the diagram.
Other potential restrictions must be considered in a similar light.

The good news is that, with due care to the calibration, the 
method
can be applied to give unbiased distances to individual galaxies in the field
as long as the inclusion of those galaxies is not restricted in
linewidth.  In other words, there will {\it not} be a correlation
between luminosity and distance within a group as found be Kraan-Korteweg et 
al. (1988) nor a correlation between H$_0$ and redshift as found by Sandage 
(1994$a$).
The method will break down if the target galaxy is a dwarf
intrinsically fainter than the limit of the calibration.  
The latter issue is only a concern in our immediate
neighborhood, not for the H$_0$ problem.

Sandage (1999) \markcite{san3} has re-articulated the proposition in Sandage 
(1994$b$) of
a 40\% bias adjustment to H$_0$ so a few words are in order.  Two points
should be appreciated.  First, bias errors are expected to depend on the 
square of the dispersion in the distance estimator relations 
(Teerikorpi 1984).  Sandage found rms scatter of $0.7^m$ with photographic
and photoelectric $B$ magnitudes from diverse sources on field samples.
With modern CCD photometry at $R$ and $I$ and better defined samples,
scatter is lowered by a factor of 2.  The resultant biases should be down 
by a factor 4.
Second, Sandage based his analysis on the `direct' luminosity--linewidth
relation.  Following Willick (1994), the bias in the `inverse' analysis 
is reduced by a factor 5.  Taking both points into account, the bias that 
Sandage describes is expected to be 
reduced by a factor 20 in the current analysis.  Even with the extreme
characterization given by Sandage our biases should be held to a few 
percent.

\section{The Template Relation at $I$ Band}
The creation of the template relation is a critical step.  In the section on
biases it has been described how important it is to have a sample that is only
limited by magnitude constraints.  Often the calibration
relationship is formed 
out of the ensemble of a field sample (Willick et al. 1996) but 
the constraints on such samples are usually ambiguous.  Also, the calibration
relationship is inevitably broadened and distorted by deviations from Hubble 
expansion 
motions.

Cluster samples have evident advantages.  It is possible to be complete 
to a magnitude limit and it can be assumed that the galaxies are 
all at the same {\it relative} distance.  The biggest concern with cluster 
samples is whether there are intrinsic differences between galaxies in a 
cluster environment and those that are more isolated.  An operational
disadvantage of cluster samples is that an individual cluster does not
provide enough systems to provide good statistics.  These two disadvantages
can be addressed simultaneously by building a template relation out of 
several cluster samples.  The `clusters' can have a sufficient range in their
properties that one can begin to evaluate the issue of environmental
dependence.  The combination of several cluster samples takes care of the
problem of poor statistics.

This study uses samples drawn from five clusters with reasonably well-known 
completion characteristics.  The nearby
Ursa Major and Fornax clusters are the best studied for present purposes.  
The completeness limits in Ursa Major are 
discussed by Tully et al. (1996) and in Fornax by Bureau, Mould, and
Staveley-Smith (1996).  After corrections for obscuration, and translation
to $I$ magnitudes, the completion limit for both clusters is $I=13.4^m$.
There are 38 galaxies in Ursa Major with type Sa or later and 
$i\ge 45^{\circ}$ brighter than this limit.  There are 16 galaxies in Fornax 
satisfying these criteria.  The $I$-band apparent magnitude--linewidth
relations for these two clusters are given in Figures~1$a$,$b$.
It was appreciated in advance
that Ursa Major and Fornax are at similar distances.  Hence the 
apparent magnitude limits conform to about the same absolute magnitude 
limits.  Fornax is indicated by these data to be $0.10^m$ closer.

Already a diverse environmental range has been explored between the Ursa Major 
and Fornax cases.  Tully at al. (1996) have labored the point that the 
Ursa Major Cluster environment is more similar to that of low density spiral
groups than to what is generally considered a cluster.  The structure must be 
dynamically young.  By contrast, Fornax has a dense core of early type
systems, evidence of a dynamically evolved structure.  Granted, the spirals
in the Fornax sample are more widely distributed than the central core and
may represent recent arrivals.

The next sample to be added to the template is drawn from the filament
that passes through what has been called the Pisces Cluster.  Aaronson et al.
(1986) and Han \& Mould (1992) have included the region in their distance 
studies but Sakai, Giovanelli, \& Wegner (1994) \markcite{sak1} have shown that 
Pisces is actually 
an extended structure with separate sub-condensations.  It is unlikely
that the region as a whole is collapsed.  Indeed, what will be considered here 
is a length of $\sim 20^{\circ}$ along the Pisces filament, which  corresponds
to an end-to-end distance of $\sim 20$~Mpc.  The mean redshift is constant to
$\sim 4\%$ along the filament though individual redshifts scatter over a 
range of $\pm 20\%$ relative to the mean.  It can
be asked if the full length of the filament is at a common distance or if
variations in distance can be identified.  A luminosity--linewidth correlation 
was constructed for the ensemble and was compared to subsets drawn from 
the separate sub-condensations.  There is no hint of any deviations 
from the sample ensemble.  Six 
sub-components along the $20^{\circ}$ filament have consistent distances to
within a few percent.  To within measurement errors, the filament is tangent 
to the plane of the sky in both real space and velocity space.

Given this circumstance, all the galaxies with $3700<V_{cmb}<5800$~km~s$^{-1}$
along the $20^{\circ}$ segment of the Pisces filament $00^h44^m<\alpha<02^h13^m$
will be taken to be at the same distance and will constitute our Pisces sample.  
Failures of this assumption can only  
act to increase the scatter of the ensemble luminosity--linewidth relationship but the 
scatter was found to be only $0.35^m$, almost as small as for any sub-component
of the template.  This scatter is obtained with 53 galaxies, after rejection of one 
object that deviates by $\sim 4\sigma$ (UGC 1416).
There is reasonable completion brighter than $I=13.8^m$ which
is taken as the magnitude limit for the present sample.  The apparent
magnitude--linewidth relation for this component is seen in Figure~1$c$.
The Pisces filament data are added to the Ursa Major/Fornax template by (1) calculating 
the distance modulus relative to Ursa Major/Fornax using the slope of the 2 cluster 
template, (2) adjusting 
for distance and redetermining a new slope now with 3 clusters, (3) iterating the 
distance offset of each cluster with the new
slope, and (4) calculating anew the 3 cluster template slope.  The distance
shifts at step 3 are of order $1\%$ and the change in slope at step 4 is 
$\sim 1\%$.

The final step in the development of the template is the addition of the
Coma and Abell~1367 clusters.  These clusters are at the same distance
to within a few percent so they were treated together until the final
iteration, at which point they were considered separately against the mean 
relation.  Only galaxies within $4.3^{\circ}$ of the cluster centers were accepted
and the velocity constraints described by Giovanelli et al. (1997$a$) are 
adopted.  As with Pisces, there is substantial but not full completion
to $I=13.8^m$.  The apparent magnitude--linewidth relations for these
clusters are shown in Figures~1$d$,$e$.
Iterations like those described above with the Pisces filament rapidly 
converged to produce the final five cluster template.  There could have
been a problem if there is curvature in the template, as might be
indicated if, say, the slope flattened for samples with more
luminous cutoffs (more distant clusters).  However there is no suggestion of
such a flattening if the luminosity-dependent extinction corrections are applied.  
The Coma sample provides 28 galaxies and Abell~1367 adds 20,
after one $5\sigma$ rejection (NGC 3832).

In total, there are 155 galaxies in the 5 cluster template after rejection of
two $>4\sigma$ outliers.  The combined
magnitude--linewidth relation is seen in Figure~2 with shifts in magnitudes
to the Fornax, Pisces, Coma, and Abell~1367 samples to bring them in line 
with the Ursa Major sample. The straight line is the least squares 
regression with uncertainties in linewidths.  While there are three
distinct absolute magnitude cutoffs (UMa/Fornax; Pisces; Coma/A1367),
the slopes are identical within the uncertainties.  
The relation is effectively linear within the absolute magnitude range we
explore and with our specifications of magnitudes and linewidths.

\section{The $B,R$ and $K^{\prime}$ Relations}

Less information is available for bands other than $I$.  However,
inter-comparisons are valuable because of the potential problem with
obscuration.  Paper~I provides data at $B$ and $R$ for the calibrators, while 
Paper~II contains data for all the galaxies in the Ursa Major sample, most of 
those in Coma, 
and most of the systems within the Pisces region at $00^h49^m<\alpha<01^h32^m$.
The $B$ and $R$ magnitude--linewidth relations are given in Figure~3 for 
the same assumed relative distances between UMa, Coma, and Pisces as 
indicated by the $I$-band correlation.
Material is available at $K^{\prime}$ for the same Ursa Major and Pisces 
galaxies (Tully et al. 1996; RSTW).  The \kp\ 
magnitude--luminosity relation for these two clusters is shown in Figure~4.

The magnitude scatter is essentially the same at $R$ and $I$ and 
$\sim 20\%$ worse at $B$ and \kp.  The $B$-band is most sensitive to variations 
in recent star formation and variations in the extinction at a given linewidth.  
Extinction corrections diminish toward the infrared until they are modest at 
$K^{\prime}$.  However, rapidly increasing sky background toward the infrared 
results in increasingly larger errors in the extrapolated total magnitudes.  For 
example, at 
$K^{\prime}$ one loses almost 2 scalelengths to the sky compared with an $R$ 
exposure of the same duration.  
The correlations are seen to steepen toward the infrared.  However, this 
steepening is less extreme than had been seen in the past (Tully, Mould, \&
Aaronson 1982) \markcite{tu9} because of the 
strong luminosity dependence of the reddening corrections that are now applied.
The biggest corrections are made to the most luminous galaxies in the bluest
bands.  Hence the corrected relations at shorter wavelengths are steepened
toward the slopes of the almost-reddening-free infrared relations.  As shown
in Tully et al. (1998), only a weak color dependency on luminosity remains 
after reddening is taken into account.  Slopes at $B,R,I$
are $-7.3$, $-7.6$, and $-7.9$, respectively, for the 91 galaxies with data in 
all these bandpasses, and the slope at \kp\ is $-8.8$ for 65 of the same galaxies.
The slope quoted in each case is based on the regression with errors in 
linewidths which is appropriate for bias-free distance determinations.
The physically meaningful slopes are flatter by $\sim 0.3$.
There appears to be
convergence in the infrared toward $L \propto V_{max}^n$ where $n=3.4 \pm 0.1$.

\section{Absolute Calibration}
The absolute calibration data are taken from Paper~I ($B,R,I$) and from RSTW 
($K^{\prime}$).  In order to be consistent with the cluster template sample 
described in section 4 we restrict the calibration sample in luminosities to 
$M_I < -17.9$ and in inclinations to $i > 45^\circ$ (2 galaxies initially
thought to have $i > 45^\circ$ but
ultimately assigned inclinations slightly below $45^\circ$ are retained).  
Currently there are 24 
galaxies which meet these criteria that have distances based on
observations of Cepheid variable stars.  Most of the observations were made
with the Hubble Space
Telescope (Freedman et al. 1997, \markcite{fr2} Sandage et al. 1996, 
\markcite{san3} Tanvir et al. 1995,  \markcite{ta} Newman et al. 1999).
\markcite{new}
In order to be consistent, whenever possible the most recent distance provided
by the HST Key Project Team is taken; ie. that given by Sakai et al. (1999).
\markcite{sak2}
This reference includes distances from the team 
reanalysis of studies first-authored by Saha, Sandage, and Tanvir, 
respectively (Gibson et al. 1999). \markcite{gib}  Sakai et al. also report on 
minor 
adjustments to the moduli of NGC's 1365, 4535, and 4725 (Ferrarese et al. 
1999).  \markcite{fer} The $B,R,I$ luminosity--linewidth
correlations are shown in Figure~5 for these 24 galaxies,
where now {\it absolute} magnitudes are plotted based on the measured 
distances.  

It is inappropriate to construct the luminosity--linewidth relation from the 
absolute calibration data alone because {\it these galaxies do not constitute a 
complete sample.}  However if the
calibrators are drawn from a similar distribution as the template objects, 
with no restriction in linewidths, then
each of the 24 galaxies with independent distances provides a 
separate zero-point calibration of the template relations.  The least-squares 
average provides the optimum 
fit and these are shown as the dashed lines in Fig.~5.  Note the remarkable
consistency.  {\it The slopes shown in Fig.~5 do not come from the absolute
calibration
data; rather they are given by the cluster templates.  With only the one degree
of freedom of the zero point, the scatter at $R$ and $I$ is only $\sim 0.24^m$.}
In Paper~I, fits were made to the calibration sample alone with essentially
identical slope and intercept determinations.  
This result strongly reinforces the hypothesis that the calibrators have similar 
properties to the cluster template galaxies.

The zero-points of 
the absolute relationships specify the distance moduli of the five
template clusters.  This information is used to superimpose
the template relations on the absolute
calibrators, as shown in Figure~6.  Panel $c$ shows the $I$ band 
luminosity--linewidth relation with
the 24 calibrators and the 155 cluster template galaxies shifted to the 
absolute magnitude scale of the calibrators.  
The $B$ and $R$ relations are shown in panels $a$ and $b$. 
In these cases 91 galaxies are available for the templates and there are
the same 24
calibrators.
Information at \kp\ is more limited but consistent.  RSTW
provide \kp\ photometry for four galaxies with Cepheid distances.
The \kp\ zero-point calibration can be seen in 
Figure~6$d$ with UMa and Pisces cluster data superimposed.

The agreement between bands is excellent.  A measure of the agreement is given
by the distance modulus in each band determined for the UMa cluster.  The 
$I$ band analysis (155 template, 24 calibrators)
gives shorter distances than the weighted mean by 0.02,
the modulus at $B$ gives longer distances by 0.04,
the modulus at $R$ is larger by 0.03 (for both bands: 91 template, 24 calibrators),
and the modulus at \kp\ is smaller by 0.05 (65 template, 4 calibrators).
To obtain completely consistent results between bands, we average over the
four bands with weights dependent on the square roots of the numbers of template
and calibrator galaxies and the squares of dispersions.  Relative weights are
$B$:$R$:$I$:\kp = 0.46:0.66:1.00:0.25.
Once these few percent corrections are 
made, the following calibrations are indicated:
\begin{equation}
M_B^{b,i,k} = -20.11 - 7.27({\rm log}W_R^i - 2.5)
\end{equation}
\begin{equation}
M_R^{b,i,k} = -21.12 - 7.65({\rm log}W_R^i - 2.5)
\end{equation}
\begin{equation}
M_I^{b,i,k} = -21.57 - 8.11({\rm log}W_R^i - 2.5)
\end{equation}
\begin{equation}
M_{K^{\prime}}^{b,i,k} = -23.17 - 8.78({\rm log}W_R^i - 2.5)
\end{equation}
The rms scatter: at $B$ $\pm 0.38$ template, $\pm 0.30$ calibrators;
at $R$ $\pm 0.34$ template, $\pm 0.25$ calibrators;
at $I$ $\pm 0.34$ template, $\pm 0.23$ calibrators;
at \kp\ $\pm 0.44$ template, $\pm 0.24$ calibrators.  
These results are consistent with those found in Papers~I and II.  
The larger template scatter at \kp\ appears to be partially due to the increased 
fractional representation of low luminosity systems.

Figure~7$a-e$ presents $I$-band material for the five clusters that 
contribute to the template.  Data for each cluster are plotted separately to 
show clearly the fits to the individual clusters.  There is no evidence contrary  
to the hypothesis of a consistent luminosity--linewidth correlation from
cluster to cluster, whatever the range of local environments.

\section{The Hubble Constant}
Now that the template relations have been converted to absolute scales, 
they can be used to determine distances to any appropriate galaxy or 
cluster sample.  It would be dangerous to extrapolate for
targets intrinsically less luminous
than $M_B^{b,i,k}=-16.6^m$, $M_R^{b,i,k}=-17.6^m$, or 
$M_I^{b,i,k}=-17.9^m$, the low-luminosity limits of the template 
relations.  If the goal is to measure H$_0$, these limits are of 
little concern because the clusters are chosen to be distant in 
order to minimize the effects of non-Hubble motions.
Existing surveys of these clusters are limited to the more 
luminous members.  

In a future paper we will apply the calibration described here and in 
Paper~I to measure distances to 
hundreds of field galaxies in order to characterize the local 
velocity field.  For the moment, with the interest of maintaining
as homogeneous a set of measurements as possible, the H$_0$ determination
will be based on the 5 clusters that went into the template plus 7 other
clusters each with of order a dozen observed members.  The photometric data 
come from Han (1992), Mathewson et al. (1992), and Giovanelli et al. 
(1997$a$).  The fits are shown to 
the 7 additional clusters in Figure~7$f-l$.  In these cases 
the samples are not complete.  It has been argued above that each distance
measurement is unbiased if the fit is done with the ``inverse'' regression, 
so the group distance moduli are given by the least 
squares minimization of the template regression on whatever information is
available in the group.  The distance moduli measured to individual galaxies
in the 12 separate clusters are shown in Figure~8.  As anticipated in the 
discussion of biases, the present analysis provides distances that are 
{\it not} dependent on magnitude.  The effects seen by Bottinelli et al. 
(1986), Kraan-Korteweg et al. (1988), and Sandage (1994$a$) are not found.
If there are tendencies for distance moduli to increase toward fainter
magnitudes in the Antlia and Cancer samples, there are the opposite tendencies
in the Coma and Pegasus samples.
One could question if the
former pair are better described by a steeper relation and the latter pair
by a flatter relation.  Bernstein et al. (1994) have suggested the Coma
relation is flatter and shows less dispersion.  However it is evident from
the series of fits shown in Fig.~7 that the data can equally well be described
by the single slope of the ensemble of the template galaxies.
Deviations are within the expectations of statistical effects.

Results are summarized
in Table~2 and Figure~9.  The table provides (col.~2) the number of 
measures in the cluster, (col.~3) the rms scatter about the template
relation, (col.~4/5) the distance modulus/distance of the cluster, 
(col.~6) the velocity of the cluster in the CMB frame as given by 
Giovanelli et al. (1997$a$), and (col.~7) the measure of H$_0$ from the 
cluster.
The velocity given to the Pisces filament is the average of the values for
the three main sub-condensations.

The error bars in Fig.~9 contain both distance and velocity components.
The errors associated with distance depend directly on the rms dispersion
in a cluster and inversely with the square root of the number of galaxies
in the cluster sample.  The error associated with velocity streaming is taken
to be 300~km~s$^{-1}$.  The velocity component to the error is dominant
inside 2000~km~s$^{-1}$.  The statistical errors in distance become the
dominant factor beyond $\sim 6000$~km~s$^{-1}$.  The symbols in Fig.~9
specify different regions of the sky (see caption).  There is a hint of 
systematic deviations: for example the filled circles (except for nearby 
Ursa Major) lie above the open circles. More data are clearly needed to 
address this possibility.  For now, the best estimate of H$_0$ is 
derived by taking an average of
log~H$_0$ values with weights proportional to the inverse square of the
error bars that are plotted.  The result is  
H$_0 = 77$~km~s$^{-1}$~Mpc$^{-1}$.  

\subsection{Evaluation of Errors}

What are the uncertainties?  The present zero-point is based on the distance
scale established by the Cepheid period-luminosity relation and the zero-point
of that relation is based on a distance modulus of the Large Magellanic
Cloud of 18.50.  The 95\% confidence accuracy of this scale is $\sim 10\%$ 
(Madore \& Freedman 1998). \markcite{ma2}  There is also debate about a 
possible metallicity effect in the Cepheid luminosities 
(Kennicutt et al. 1998). \markcite{ke2} Almost all the calibrators used 
here are more metal rich than the LMC.  This possibility would lead to a 
correction of $\sim5\%$ in the sense that H$_0$ would be reduced.
An additional source of systematic error may arise from charge transfer 
effects within WFPC2 on HST (Stetson, private communication).  Subsequent
potential errors in color are amplified in the extinction-corrected 
Cepheid distance. Corrections could 
act to decrease the distances to those calibrators 
observed with HST and thereby increase the derived value of H$_0$ 
by $\sim 5\%$.
There are problems in the reliance on the LMC because, not only is there
uncertainty in its distance, but it is not the same kind of galaxy that
otherwise interests us and the Cepheids are observed in the LMC with different 
instrumentation.
Arguably a better alternative is to use NGC~4258 as the fundamental calibrator.
A distant accurate to 7\% is inferred from the geometry and motion of 
circum-nuclear masers (Herrnstein et al. 1999). \markcite{he} 
This distance measurement bypasses
the many steps of the distance ladder approach and the claimed accuracy is
comparable to that touted for the LMC.  NGC~4258 is a normal galaxy in our
sample and the Cepheid population has been studied with HST 
(Maoz et al. 1999). \markcite{mao}  The maser distance modulus of 29.29 
differs from the 
Cepheid modulus based on the LMC calibration of 29.54 by $-$0.25~mag.
{\it If the scale established by the maser observations is used in 
preference to the LMC calibration then all moduli are reduced by $0.25^m$
and H$_0$ is increased by 12\% from 77 to 86 \kms\ Mpc$^{-1}$.}  It could be
argued that the NGC~4258 calibration should be used in preference to the LMC 
calibration, or that one should average to get an intermediate result.
We continue to use the LMC calibration simply to make comparisons easier
with other work.
All the uncertainties mentioned in this paragraph are intrinsic to the 
zero point and common to all but a few methods of determining extragalactic
distances.
These uncertainties at the level of $10-15\%$ are not included in the errors
we quote that are related only to the
current analysis.

Our biggest source of statistical uncertainty remains the zero-point 
calibration.  The present discussion concerns the $B,R,I$ data.  The 
considerably more uncertain \kp\ calibration is discussed by RSTW. 
The fits illustrated in Fig.~5 are evaluated by tests involving
the following reduced $\chi^2$ parameter:
\begin{equation}
\chi^2 = {\sum_i^N ((M_i-M_{0,i})/\sigma_{cal})^2 \over (N-1)}
\end{equation}
where $M_i$ is the absolute magnitude of the $i^{th}$ of $N=24$ calibrators 
and
$M_{0,i}=a+b({\rm log}W_{R,i}^i-2.5)$ is the expectation magnitude.
The slope, $b$, is fixed to the template values of $-7.27$ at $B$, $-7.65$ at
$R$, and $-8.11$ at $I$ and the zero-point, $a$ is varied.  The dispersion,
$\sigma_{cal}=0.24$, is taken from the fits shown in Fig.~5.  The variation of
$\chi^2$ with change of the zero-point is shown in Figure~10$a$.  Then the 
linkage between the dependence on zero-point (the value of the correlation at 
${\rm log}W_R^i=2.5$) and the inferred value of the Hubble Constant is 
shown in Figure~10$b$.  The value at the minimum is normalized by the fit 
shown in Fig.~9.  In $I$ band, the best case, the 95\% probability level 
corresponds to an uncertainty of $\pm 6$ km s$^{-1}$ Mpc$^{-1}$ in the 
Hubble parameter ($\pm 8\%$).  The $\chi^2$ tests at $B,R$ and
\kp\ are also shown.

The statistical uncertainties associated with the fits to the 12 clusters seen 
in Fig.~9 are somewhat smaller.  The following $\chi^2$ evaluator was
considered:
\begin{equation}
\chi^2 = {\sum_i^N w_i (({\rm log H}_i - {\rm log H}_0)/\sigma_{clust})^2
\over \sum_i^N w_i} 
\end{equation}
where ${\rm log H}_i$ is the measure of H$_0$ from the $i^{th}$ cluster,
the weight is $w_i$, and $\sigma_{clust} = 0.03$ is a typical value for the
error bars in Fig.~9.  The variation of $\chi^2$ with ${\rm log H}_0$
is shown in Figure~11.  The 95\% probability constraints are 
$\pm 5$ km s$^{-1}$ Mpc$^{-1}$
in the Hubble parameter ($\pm 6\%$).

The information available from the $B,R$ and \kp\ relations is currently 
limited to only 2 clusters outside the Local Supercluster (1 at \kp) but is 
consistent with the $I$ distance measures at the level of 2\%. 
The \kp\
analysis by RSTW lead to a value of 
H$_0 = 81$~\kms Mpc$^{-1}$.  The 5\% larger value comes almost entirely from
use of only 1 cluster beyond 
the Local Supercluster rather than 10 in the 
case of the I-band.  If the analysis at optical bands is 
restricted to exactly the same sample as that used at \kp\ then H$_0=80$
is found, essentially the same as the \kp\ result.

Color plots provide a check of possible random or systematic errors in the 
data (see also Papers~I and II).  Figure~12 compares $I$ band 
results with those at $B,R$, and \kp.
The $B-I$ panel contains a hint of a systematic effect, at the level of
$0.1^m$, with Ursa Major galaxies redder than the mean and Pisces galaxies
bluer than the mean.  The Coma and calibrator galaxies are consistent with
the mean.  The Ursa Major and Pisces deviations have significances of
$\sim 3\sigma$.  The same effect is seen at $R-I$, with similar 
significances, though now the offsets are only at the level of $0.05^m$.
The offsets remain marginally significant because the overall
color--magnitude correlation is so tight at $R-I$.  At $I-$\kp\ the offsets
between subsamples disappear.  These small systematics could be explained 
if Galactic reddening is underestimated in the Ursa Major region and
overestimated in the Pisces region by $E(B-V)\sim0.04$ which is larger
by a factor of 2 than the expected uncertainty.  Another possibility 
is that there are systematic variations in the star formation histories 
of the various samples, but for the measurement of H$_0$ the problem is 
minor.

The measurement of inclinations (see section 2.2) remains one of the 
more problematic issues.  However, the magnitude residuals from the $I$ band
correlation shows no hint of dependency on inclination (see Figure~13).
Substantially different extinction corrections produce results 
which agree at the level of 0.5\% (section 2.4).  It is particularly 
comforting that the \kp\ results are consistent with the other bands 
since extinction issues are unimportant at that wavelength.

Galaxies of type Sa tend to lie below the mean relations (cf, 
Rubin et al. 1985, Pierce \& Tully 1988; Verheijen 1997 \markcite{ve} discusses 
the issue in
terms of the forms of rotation curves) and the small effect is seen in the
plot of residuals versus type in Figure~14.  See the small symbols in 
Fig.~7.  The derived distances for the Sa systems are larger
than the mean (see Fig.~8).  This class is sufficiently few in number that
the problem is ignored in the present analysis.
The luminosity--linewidth relations are found to be consistent in 
environments as diverse as the Coma Cluster and the Pisces filament or the
spiral-rich Ursa Major Cluster.  Although environmental dependencies are
possible, there is no evidence of any such effect.

Excluding uncertainty in the Cepheid scale,
statistical errors added in quadrature amount to $\pm 8$ units of H$_0$ 
(95\% confidence).
The largest statistical error is still in the fit to the zero-point 
calibration.  In Paper II we show that the residuals in the different 
bandpasses are highly correlated.  Since the measurement errors 
in linewidth are small, it is implied that the scatter in the 
luminosity--linewidth correlation is either intrinsic or dominated by inclination 
corrections, particularly to the linewidths.  Bothun \& Mould (1987) \markcite{bo}
and more recently Giovanelli et al. (1997$b$) provide a detailed description of 
the components of scatter in the luminosity--linewidth relation.  Arguably
the most intractable problem in terms of further improvements is in 
inclination measurements/corrections.

The current determination of H$_0$ is lower than in earlier days with the 
same methodology (eg, H$_0=85$, Pierce \& Tully 1988) and the primary reason
is seen in Figure~15.  There were only 5 calibrators available before the 
launch of Hubble Space Telescope and {\it those 5 are seen to deviate by 0.31~mag
in the mean} (a 16\% effect on distances).  NGC~3031 has subsequently been 
reobserved with HST (Freedman et al. 1994) \markcite{fr1} though the difference 
with the
ground based result (Madore et al. 1993) \markcite{ma3} is tiny.  
The remaining uncertainty in the charge transfer efficiency of the HST/WFPC2 
CCDs could 
conceivably account for some of this offset, say up to 6\%, but we 
have no reason to think that most of the effect is anything more 
than a statistical fluke.  

\section{Comparison with Literature Results}

The turbulent history of Hubble Constant measurements has received its share
of attention.
If we restrict this discussion to just luminosity--linewidth
determinations of H$_0$ and just those that have benefited from HST 
Cepheid calibrations, there is still the following remarkable range of
results (all errors quoted from the original sources and usually $1\sigma$): 
Theureau et al. (1997) give H$_0=53\pm5$; Ekholm et al. (1999) \markcite{ek} give
H$_0=53_{-5}^{+6}$; Federspiel, Tammann, \& Sandage (1998) \markcite{fed} give 
H$_0=57\pm7$;
Watanabe, Ichikawa, \& Okamura (1998) give H$_0=65\pm2_{-14}^{+20}$;
Shanks (1997) \markcite{sha} gives H$_0=69\pm8$; Giovanelli et al. (1997$c$) 
\markcite{gi2} give H$_0=69\pm5$;
Sakai et al. (1999) give H$_0=71\pm4\pm7$; and this paper gives H$_0=77\pm8$
(95\%).

The straight average of these values is H$_0=64\pm9$.  Why should we have much
confidence in our lonely value on the high side?  Several of the above-mentioned
papers carried out quite sophisticated analyses.  However there is no adequate
substitute for good data.  We will try to argue that our data are generally 
superior.

Consider the lowest estimates first.  Both Theureau et al. (1997) and
Ekholm et al. (1999) find H$_0=53$.  These collaborators draw data from the
Lyon-Meudon Extragalactic Database, the source and extension of the Third
Reference Catalogue (de Vaucouleurs et al. 1991). \markcite{de}  Magnitudes 
are at $B$ from 
many photographic or aperture photoelectric references, inclinations are mostly
from the photographic sky surveys, and linewidths are from many sources and 
mixed quality.  Several thousand galaxies are considered.  Template relations 
are derived from field populations, broadened and distorted by velocity
streaming.  Theureau et al. work with the so-called 
`direct' luminosity--linewidth relation, the correlation with errors in
luminosity, and must deal with the full brunt of the 
Malmquist magnitude limit bias.  Since the scatter is large with 
their mixed-bag data set, there 
is the potential for a large bias.  One possible signature of a strong bias 
is an upturn in the measured H$_0$ at larger distances.  
However an upturn in 
H$_0$ {\it could} alternatively arise as a result of local structure;
certainly plausible since we live in a filament and the gravity of the
filament could be slowing the local expansion (e.g. Tully 1988).  
Theureau et al. (1997) follow the procedures of Bottinelli et al. 
(1986) who interpreted the upturn in H$_0$ as
due to bias, not gravity, with the consequence that Bottinelli et 
al. inferred a low H$_0$.  The Theureau et al. analysis is not as
transparent but appears to rely on the same underlying assumption that 
there is a local regime where expansion actually represents the 
universal value and that the data sampling larger scales are
biased.  The fundamental problem is that their analysis procedure (use of 
the direct relation) requires bias corrections but the amplitude of the
corrections can easily be disputed.

Ekholm et al. (1999) work with the `inverse' luminosity--linewidth relation,
as we advocate.  They found H$_0=72$ for the $B$ band relation (and H$_0=78$
if diameters are substituted for luminosity).  How could that result be
reconciled with H$_0=53$ from the direct relation?  Ekholm et al. argue 
that reconciliation is possible {\it if they assume that the Cepheid 
calibrators follow a relation with a different slope from that of
galaxies in the field}.
Their justification for the slope difference is largely based on the 
aberrant location of {\it one} calibrator galaxy which is fainter than 
almost all $\sim 2000$ galaxies in their field sample (unidentified but 
probably NGC~3109).  
They argue that the different slope introduces a bias of 25\% 
which they can correct for and assure us that their overall 
$1\sigma$ error budget is 10\%.  We would argue
that if the calibrator and field slopes are different then the relations 
are different and just about anything is possible.  Fortunately, with our 
own photometry there is no evidence that calibrators with known distances 
and other targets are 
drawn from different relations.

Federspiel et al. (1998) derive H$_0=57$ from a distance they measure to the 
Virgo Cluster and a comparison with more distant clusters out as far as
11,000~\kms.  The comparison with distant clusters uses information from
Jerjen \& Tammann (1993) \markcite{jer} that involves other input than the 
luminosity--linewidth method so we restrict ourselves here to the issue of 
the  Virgo Cluster distance.  Federspiel et al. also draw data from the same 
source as Theureau et al. and Ekholm et al. so it is data with a fair
amount of scatter.  Depth effects are a particular problem with the 
Virgo Cluster (Pierce \& Tully 1988; Yasuda et al. 1997). \markcite{ya}  
The spiral population 
is probably experiencing substantial infall (Tully \& Shaya 1984).  
\markcite{tu6} 
There is evidence of considerable contamination from infalling groups on
the far side of Virgo, projected onto the cluster and therefore 
indistinguishable in velocity.  There are presently 7 galaxies in the 
Virgo Cluster region with Cepheid distance measurements.  
The individual moduli are 30.87, 30.95, 31.03, 31.04,
31.04, 31.10, and 31.80.  The distance obtained for NGC~4639 is 
strongly deviant.  The average of the first 6 (excluding NGC~4639) 
is $31.01\pm0.08$.  The deviation of NGC~4639 is 0.79~mag which 
puts it 7~Mpc in the background.
It is at least plausible, if not probable, that NGC~4639 is in the 
background at the approximate distance of the so-called Virgo~W$^{\prime}$
structure.  If the true Virgo Cluster is at 
16~Mpc as determined by the 6 galaxies with Cepheid distances excluding 
NGC~4639 then, all other other steps preserved, Federspiel et al. would
have found H$_0=74$.  The merits of this background issue can be debated,
but the Federspiel et al. value of H$_0$ rests on this tenuous point.
Sandage (1994$b$) has found H$_0=48\pm5$ with the luminosity--linewidth
method applied to field samples.  That study predates the availability of 
HST Cepheid distances so will not be discussed further than to recall the
discussion in section 3 and to say our additional
criticisms would resemble those brought up in the Theureau et al. discussion.

Watanabe et al. (1998) find H$_0=65$.  With the revised calibrator distances
of Sakai et al. (1999) they would have gotten 67.
The photometry for their 
field samples comes from photographic material at $B$-band but 
at least the source is homogeneous and
is calibrated with CCD imaging.  The source for HI linewidths is also 
homogeneous.  The 
maximum likelihood analysis has merit.  Watanabe et al. had 10 calibrators
with Cepheid distances available.  If they used all 10 and converted to the
Sakai et al. distance revisions then they would have obtained
H$_0=69$.  They prefer not to use 5 of these
calibrators because of issues having to do with the detailed 
velocity fields or
inclination ambiguities.  They reject half their calibrators on criteria
that are not applied to their field sample.  We use the 
calibrators they reject, 
so we feel the appropriate Watanabe et al. result to compare with our 
own is H$_0=69$.
Even so, the Watanabe et al. result is 10\% lower than our value,
barely within our 95\% uncertainty.  One 
possibility for the remaining discrepancy might lie in the use of 
heterogeneous data for the calibration sample.  
Are the magnitude, inclination, and linewidth
parameters that they take from the literature really all 
on the same system as their field sample?  

As an aside, note that we almost never reject candidates that satisfy
magnitude and inclination constraints and are typed later than Sa.
In a small number of cases 
extremely pathological or interacting galaxies are rejected a priori.
In the current template plus calibrator sample, only 2 of 181 initial
candidates
were rejected because of $>4\sigma$ deviance.  All available galaxies
with Cepheid distances are used.

Shanks (1997) finds H$_0=69$.  
To facilitate a comparison with the present work,
consider only the luminosity--linewidth analysis by Shanks (ie, not the
analysis based on SN~I$a$ distances), reject NGC~4496A not used in the 
current paper because of its face-on inclination, and update the Cepheid 
distances used to those in the current paper (including those by Gibson 
et al. 1999).
With these changes, Shanks would have found H$_0=73$, within
$1\sigma$ of our present result.

Giovanelli et al. (1997$c$) also get H$_0=69$.  Six of their 12 calibrators 
have revised distances in Sakai et al. (1999), whence  H$_0=70$.
This result is only 
marginally consistent with what we find in spite of a big overlap
in data.  We made a concerted effort to track down the systematic difference
with only partial success.  There is good consistency between our raw
magnitudes, inclinations and linewidths.  A problem arises with extinction
corrections to magnitudes (discussed in section 2.4) but the analyses
were carried out with our alternate prescriptions and each produces the
same results if carried out consistently across templates and calibrators.  
A nuance of the bias problem does generate a 5\% effect that,
if our viewpoint is accepted, takes Giovanelli et al. to H$_0=73$.
Those authors use a maximum likelihood fitting procedure that requires
bias corrections, unlike our procedure.  However, upon fitting to
the zero-point calibrators with their procedure it is necessary 
to again account for biases because
lower luminosity galaxies are badly under represented among the calibrators.
However, in this case the sense of the correction now has the 
opposite sign, increasing the Hubble Constant.  Evidently Giovanelli
et al. did not make this correction.  
A $\Delta {\rm H}_0=4$ km s$^{-1}$ Mpc$^{-1}$ difference remains 
between us.  This difference
would not seem bad except our data sets have so much overlap that the origins
of this 5\% offset should be evident.  A mystery remains.

Finally, the HST Key Project Team (Sakai et al. 1999) report H$_0=71$.
They use three major data sets.  Their acknowledged best data set is at $I$ 
band and
has a large overlap with the material used by Giovanelli et al. (1997$c$)
and ourselves.  From this material alone, Sakai et al. find H$_0=74$.
A second smaller data set at $B$ and $V$ bands gives Sakai et al. H$_0=70$.
The historic Aaronson et al. (1982, \markcite{aa3} 1986) data sets using 
aperture $H_{-0.5}$ band photometry gives H$_0=67$.  
At $I$ band where the data are best and the data overlap
is substantial, there is good agreement between Sakai et al. and ourselves.
The agreement is marginal with their $B,V$ results and poor with their
$H_{-0.5}$ results.  The situation is perplexing.  Sakai et al. find an indication
of a problem with
$I-H_{-0.5}$ color differences in the unphysical sense that their calibrators
are {\it redder} than their cluster galaxies. 
At least in our analysis 
there is consistency between results at $B,R,I$, and \kp.

The results presented in this paper are at least marginally consistent with
the other studies that use high quality data.  Once we are on the same page with
assumed distances to calibrators, the Watanabe et al., Shanks, Giovanelli et al.,
and Sakai et al. results are lower by $1-2\sigma$.
There are several features 
of the present study that arguably make it better than others.  For one 
thing, the photometric material on the local calibrators that is introduced
in Paper~I was obtained with fields sufficiently larger than the target galaxies 
that they give good sky definition.  Second, we observed near, big galaxies and 
distant,
small galaxies with common filters and procedures and have substantial sample
overlaps with other programs so there should be a reliable bridge between near
and distant.  Third and foremost, only in this study is there
serious consideration given to the issue of magnitude completion in the 
template construction.  Only if this issue is properly addressed can one
then use the `inverse' luminosity--linewidth relations with nulled biases.
Both Giovanelli et al. and Sakai et al. have constraints on linewidths that
make corrections for biases difficult.  It is this simplification of the
bias analysis through attention to the template calibration that, we claim,
gives us an advantage over the other teams with comparably good data.

{\it It is most remarkable that the slopes derived from the template relations
when slid to fit the Cepheid calibrators, with only freedom in the zero point, 
result in $R$ and $I$ scatter of only $0.24^m$.}

Endless comparisons could be made with other techniques used to derive
H$_0$.  If one asks which is the single result that causes us the most 
concern, it would be the determination of H$_0$ with supernova of type~I$a$.
For example, Riess, Press, \& Kirshner (1996) find H$_0 = 64 \pm 6$
($2\sigma$ uncertainty).  However the Cepheid distances to the SN~I$a$ host 
galaxies that were available to Riess et al. (NGC~4536: Saha et al. 1996;
\markcite{sah} NGC~4639: Sandage et al. 1996; NGC~5253: Sandage et al. 1994)
\markcite{san} have now been re-evaluated by the HST Key Project Team  
(Gibson et al. 1999).  {\it If only we use the new Cepheid distances given
by the Key Project Team in preference to the distances given by Saha and 
Sandage then the Riess et al. value for H$_0$ is increased from 64 to 73.}
The Key Project Team do a more complete calibration of the SN~I$a$ 
procedure with 7 calibrating galaxies and find 
H$_0 = 68 \pm 4$
($2\sigma$ statistical uncertainty).  At least our error bars overlap with
the SN~I$a$ results.

No single-point failure of the luminosity--linewidth
analysis is likely to produce a {\it systematic} error greater than 5\%.
Conspiratorial addition of several independent systematic errors is possible.
Uncertainties in the Cepheid calibration (LMC distance, metallicity
effects, reddening, crowded field photometry, LMC calibration relation) 
are another matter and we take what we are given.  Replacing the LMC
calibration with the NGC~4258 calibration gives a 12\% larger value of
H$_0=86$.
It is possible that the 
value of the Hubble parameter determined nearby may not reflect the 
cosmological value, a manifestation  
of the local velocity anomaly problem on a larger scale.  For example, 
Zehavi et al. (1999) \markcite{ze} raise the possibility that we live in 
an underdense
region that extends out to $\sim 6000$~\kms, whence H$_0$ would be perhaps
6\% lower than the locally measured value.  Such a possibility can be 
entertained though there is no evidence for a dependence of H$_0$ with 
distance out to $\sim 19,000$~\kms\ from the luminosity--linewidth analysis
of Giovanelli et al. (1999). \markcite{gi1}
To conclude, this study determines a value of the Hubble Constant of
H$_0=77 \pm 8$ \kms~Mpc$^{-1}$.  The error is the 95\% probable
{\it statistical} error.  Cummulative systematic errors within the present
analysis could amount to as much as 10\%.  The zero point external to this
analysis is still uncertain by $\sim 10\%$.

The only substantial change in the last decade in the measure of H$_0$ 
from 
the luminosity--linewidth method has resulted from the five times larger number
of zero-point calibrators.  Curiously, the galaxies with 
Cepheids observed from the ground lie $0.3^m$ fainter than the ensemble mean 
dominated by the galaxies observed with HST.  
Perhaps this difference is only a statistical fluke.  

\section{Acknowledgments}
Jo Ann Eder processed Arecibo HI observations that were made at our request of
several galaxies in the Coma Cluster.  Other unpublished HI line profiles were 
made available to us by Martha Haynes, Barbara Williams, and Wolfram
Freudling.

\begin{table}[htb]
\begin{center}
\caption{Data for 24 Calibrators and 155 Galaxies in 12 Clusters}
\begin{tabular}{l}
\hline
\hline
\end{tabular}
\end{center}
\end{table}

\begin{table}[htb]
\begin{center}
\caption{Five Template Clusters and Seven More}
\bigskip

\begin{tabular}{lrccccc}
\hline
\hline
Cluster & No. & RMS & Modulus & Distance & $V_{cmb}$ & H$_0$ \\
        &    & (mag) & (mag)  &  (Mpc)   & (km/s)  & (km/s/Mpc) \\
\hline
Fornax          & 16 & 0.50 & 31.25 & 17.8 &  1321  & 74\\
Ursa Major      & 38 & 0.40 & 31.35 & 18.6 &  1101  & 59\\
Pisces Filament & 53 & 0.35 & 33.90 & 60.3 & (4779) & 79\\
Coma            & 28 & 0.34 & 34.68 & 86.3 &  7185  & 83\\
Abell 1367      & 20 & 0.36 & 34.71 & 87.5 &  6735  & 77\\
\hline
Antlia          & 11 & 0.27 & 32.79 & 36.1 &  3120  & 86\\
Centaurus 30    & 13 & 0.60 & 33.02 & 40.2 &  3322  & 83\\
Pegasus         & 12 & 0.40 & 33.30 & 45.7 &  3519  & 77\\
Hydra I         & 12 & 0.36 & 33.81 & 57.8 &  4075  & 70\\
Cancer          & 15 & 0.38 & 33.96 & 61.9 &  4939  & 80\\
Abell 400       &  7 & 0.19 & 34.81 & 91.6 &  6934  & 76\\
Abell 2634      & 16 & 0.36 & 35.23 & 111.2 & 7776  & 70\\
\hline
{\bf Weighted average} & &  &       &      &        & {\bf 77}\\
\hline
\end{tabular}
\end{center}
\end{table}

\clearpage
\clearpage

\begin{figure}
\vspace{60mm}
\includegraphics{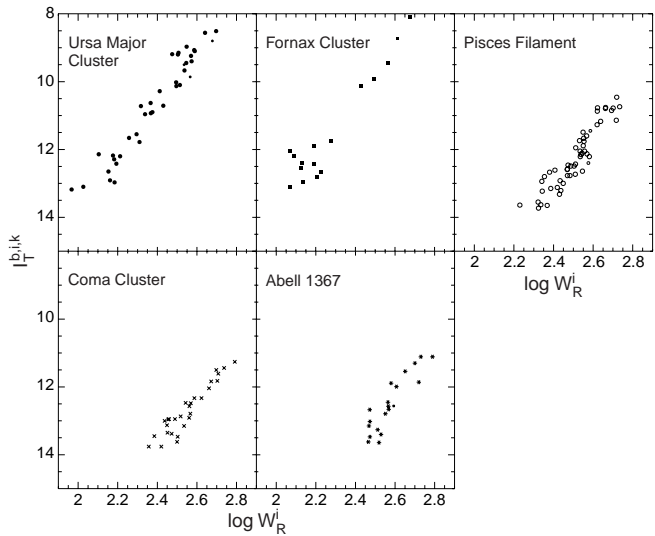}
\caption{
$I$-band apparent magnitude--HI profile linewidth plots for the five clusters 
that contribute to the template luminosity--linewidth correlations. 
Magnitudes are adjusted for internal and Galactic absorption and small
redshift corrections.
Large symbols: types Sab and later.  Small symbols: type Sa.
The Ursa Major and Fornax samples are complete to $I_T^{b,i,k}=13.4$.
The Coma, Abell 1367, and Pisces filament samples are nearly complete to
$I_T^{b,i,k}=13.8$.  Galaxies fainter than these limits are excluded.
}\label{1}
\end{figure}

\begin{figure}
\vspace{60mm}
\includegraphics{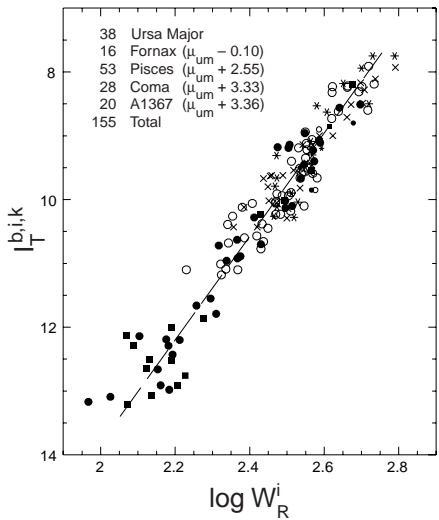}
\caption{
$I$-band apparent magnitude--linewidth relations for 5 clusters translated to 
the Ursa Major relation.  Symbols as in Fig.~1.  The straight line is a
least squares fit to the ensemble with errors in linewidths after the
iterations described in the text. 
}\label{2}
\end{figure}

\begin{figure}
\vspace{60mm}
\includegraphics{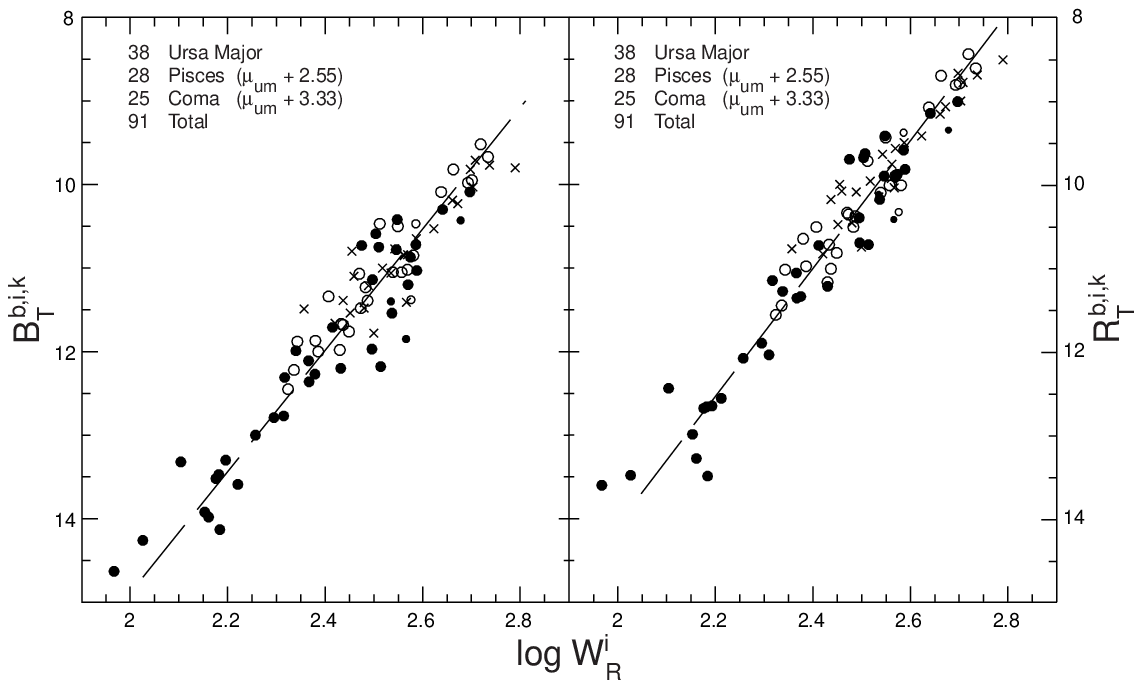}
\caption{
$B$ and $R$ apparent magnitude--linewidth relations for 3 clusters translated 
to the Ursa Major relation with the same relative distances assumed with the
$I$-band material.  Symbols as in Figs.~1 and 2.  The straight lines 
are least squares fits to the ensemble with errors in linewidths.
}\label{3}
\end{figure}

\begin{figure}
\vspace{60mm}
\includegraphics{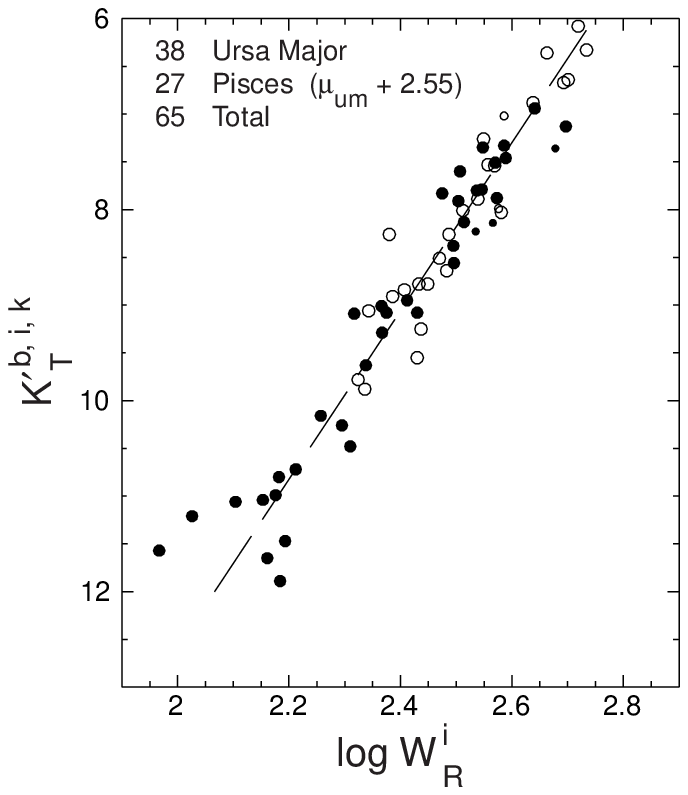}
\caption{
\kp\  apparent magnitude--linewidth relations for 2 clusters, with Pisces 
translated
to the Ursa Major relation with the same relative distances assumed with the
$I$-band material.  Symbols as in Figs.~1 and 2.  The straight line is a
least squares fits to the ensemble with errors in linewidths.
}\label{4}
\end{figure}

\begin{figure}
\vspace{150mm}
\includegraphics{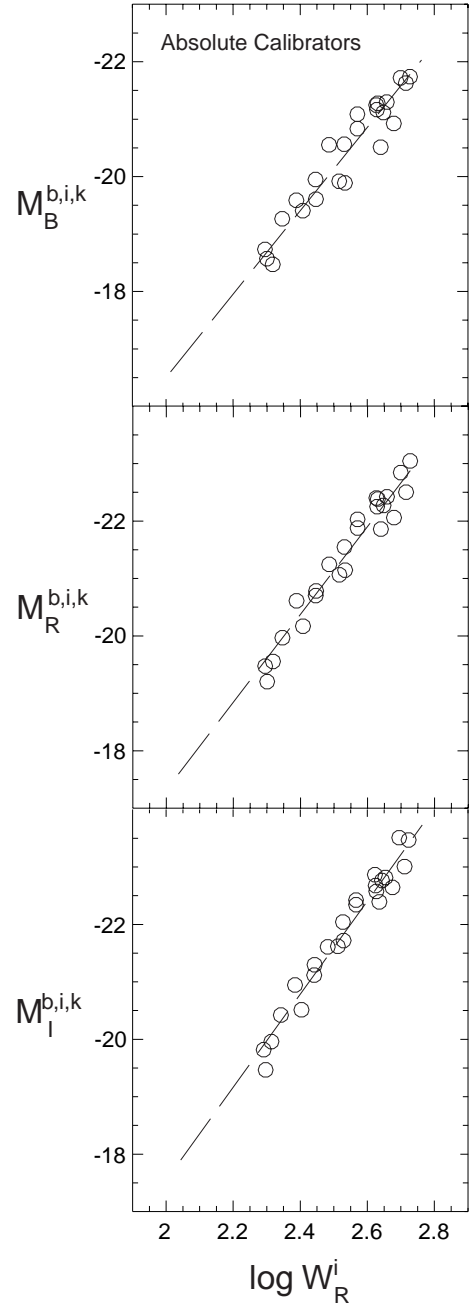}
\caption{
$B,R,I$ absolute magnitude--linewidth relations for 24 galaxies with 
independently determined distances from
application of the Cepheid period--luminosity relation.
The straight lines are the least squares best fits of the
lines shown in Figs.~2 and 3.
}\label{5}
\end{figure}
\clearpage

\begin{figure}
\vspace{160mm}
\includegraphics{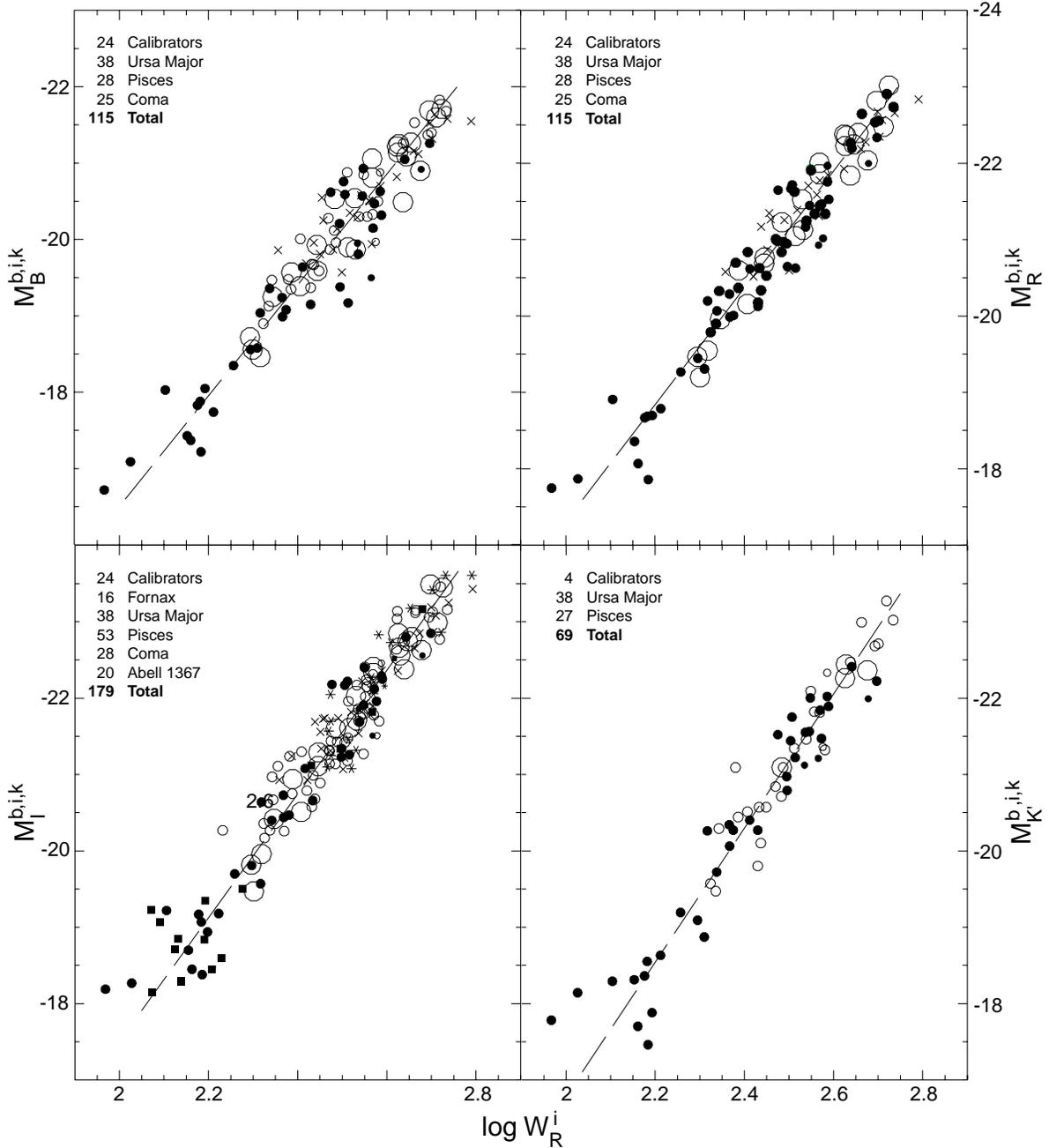}
\caption{
$B,R,I$, and \kp\ absolute magnitude--linewidth relations for the cluster 
template
galaxies translated to overlay on the zero-point calibrator galaxies.
Symbols and straight line fits as in previous plots.  The $I$ relation
involves 5 clusters and 24 zero-point calibrators, the $B$ and $R$
relations are built with 3 clusters and 24 zero-point calibrators,
and the \kp\ relation is based on 2 clusters and 4 zero-point calibrators.
Relative distances between clusters and with respect to the calibrators
are the same on all plots.
}\label{6}
\end{figure}

\begin{figure}
\vspace{160mm}
\includegraphics{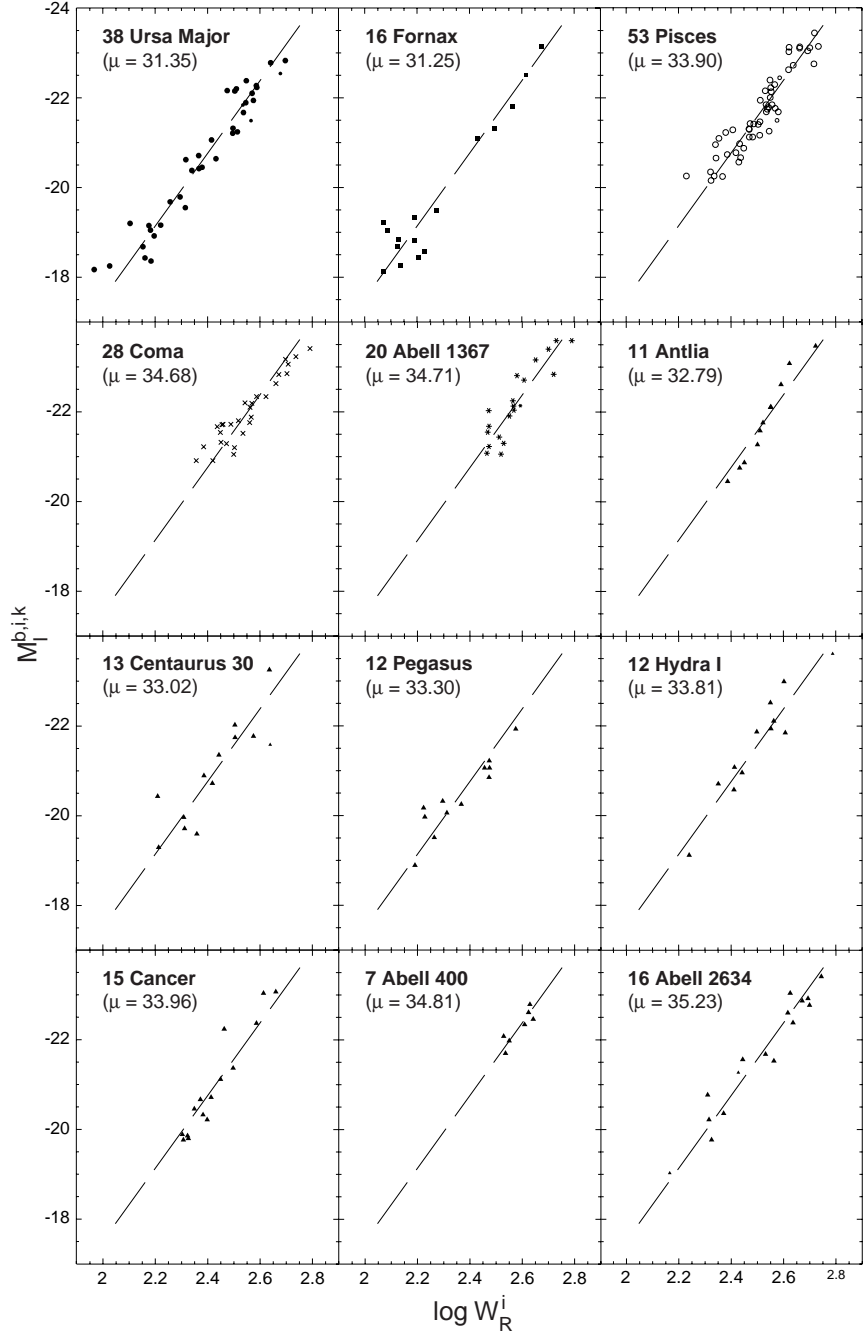}
\caption{
The $I$-band  absolute magnitude--linewidth fits are shown separately for
each of the clusters used in this study.  The first 5 (panels $a$-$e$)
contribute to the template relation.  The remaining 7 (panels $f$-$l$:
individual galaxies located by triangles)
are given distances that follow from least squares fits to the template.  
}\label{7}
\end{figure}

\begin{figure}
\vspace{100mm}
\includegraphics{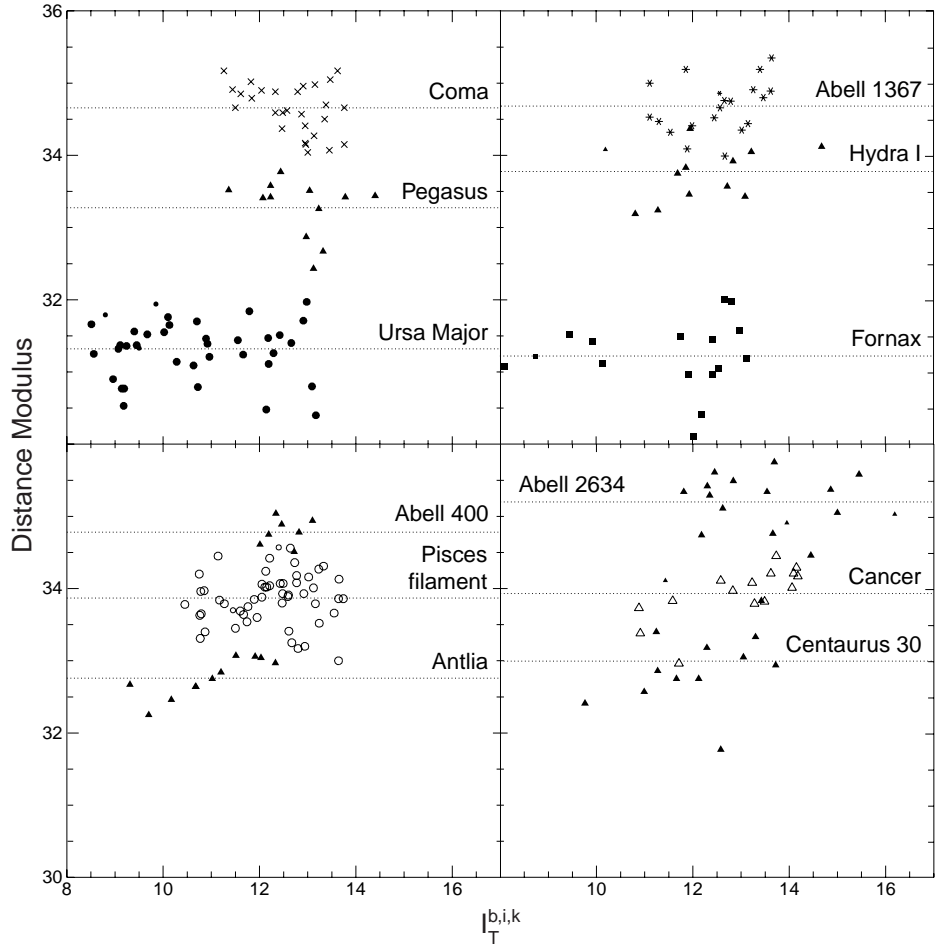}
\caption{
Distance moduli of individual cluster galaxies versus $I$ magnitude.
Symbols as in previous figures.
}\label{8}
\end{figure}
\clearpage

\begin{figure}
\vspace{75mm}
\includegraphics{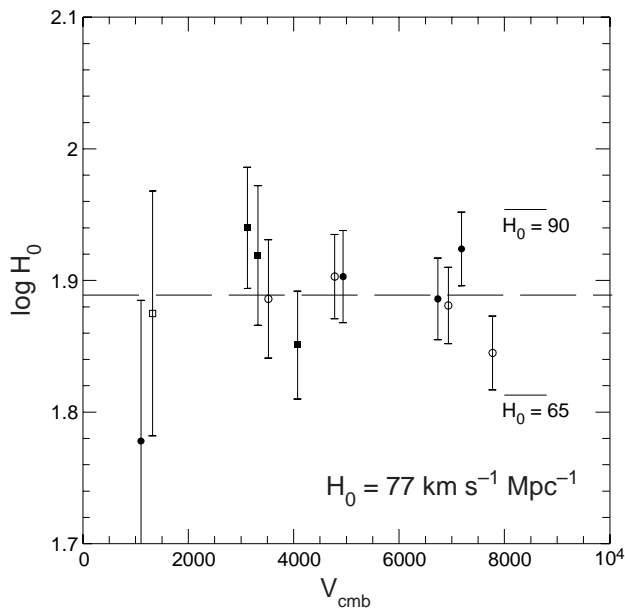}
\caption{
Values for the Hubble parameter determined for each of the 12 clusters
plotted against the cluster velocity in the microwave background frame.
Errors are a convolution of the statistical errors in distance and an
uncertainty of 300~\kms\ in velocities.
Symbols depend on location in the sky: {\it filled circles:} north celestial
and north galactic, {\it open circles:} north celestial and south galactic,
{\it filled squares:} south celestial and north galactic, {\it open squares:}
south celestial and south galactic.
}\label{9}
\end{figure}

\begin{figure}
\vspace{55mm}
\includegraphics{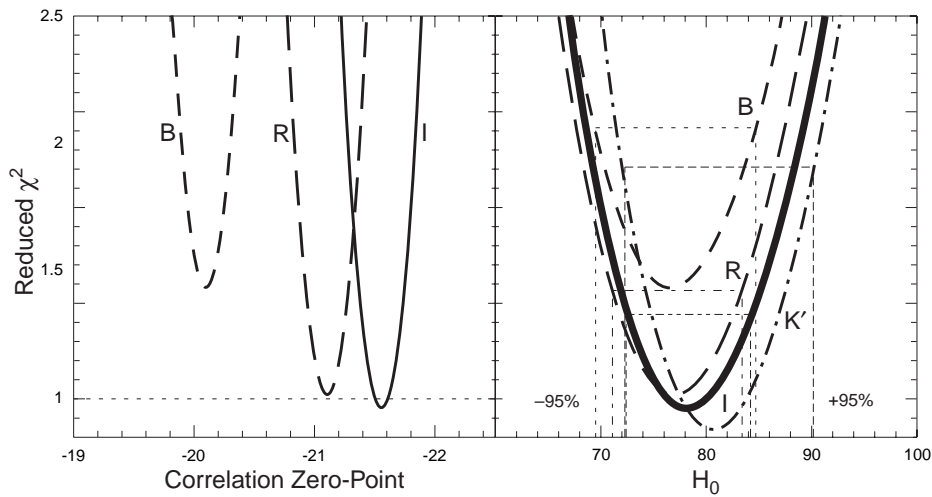}
\caption{
Reduced $\chi^2$ goodness-of-fits tests.  In panel~$a$, $\chi^2$ values are
recorded as the cluster template
relations are translated in zero-point.  In panel~$b$, the sensitivity
of $\chi^2$ on the choice of H$_0$ is shown.  The heavy solid curve is for 
the best case $I$
band data.  The curve for the \kp\ data is carried over from RSTW.
}\label{10}
\end{figure}

\begin{figure}
\vspace{60mm}
\includegraphics{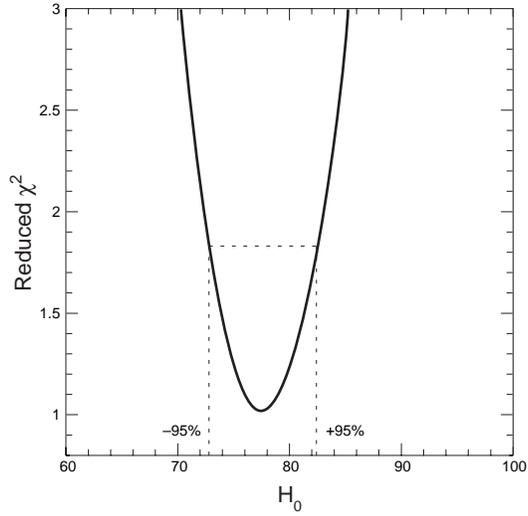}
\caption{
The $\chi^2$ dependency of the fit to 12 clusters illustrated in Fig.~8.
Weights are a function of the error bars in that figure.
}\label{11}
\end{figure}

\begin{figure}
\vspace{60mm}
\includegraphics{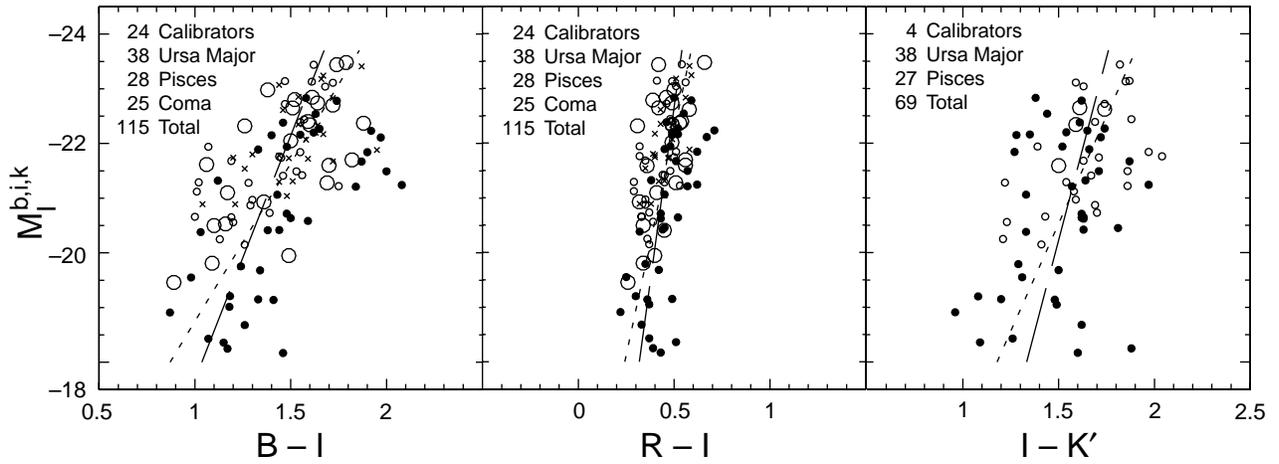}
\caption{Color--magnitude diagrams.  {\it Left:} $B-I$. {\it Center:} $R-I$.
{\it Right:} $I-$\kp.  Symbols as in previous plots.  Long dashed lines are
the regressions with errors in colors.  Dotted lines are the double 
regressions.  
} \label{12}
\end{figure}

\begin{figure}
\vspace{60mm}
\includegraphics{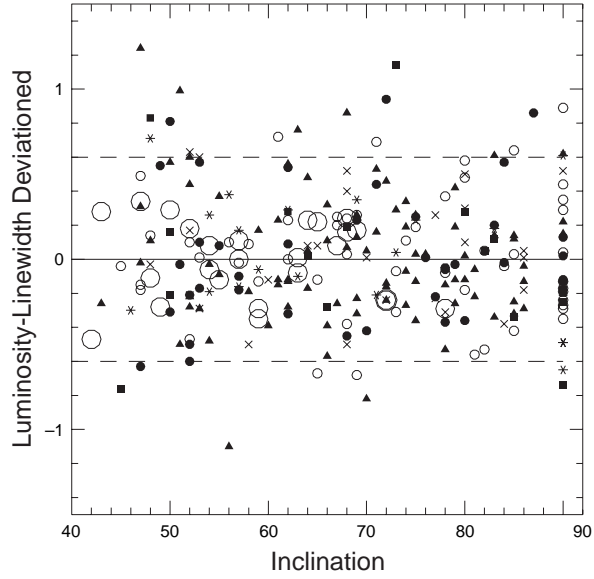}
\caption{Deviations from the $I$ band luminosity--linewidth correlation as
a function of inclination.  Symbols have the same meanings as in Figs.~6 and
7.  The dashed lines indicate the $2\sigma$ deviation boundaries.  
} \label{13}
\end{figure}

\begin{figure}
\vspace{60mm}
\includegraphics{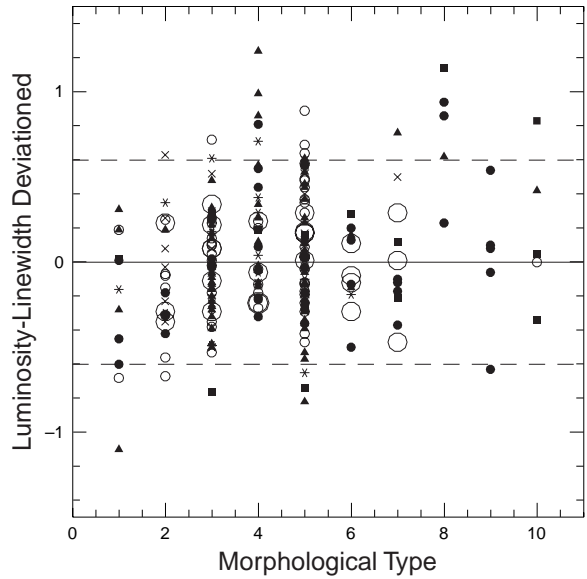}
\caption{Deviations from the $I$ band luminosity--linewidth correlation as
a function of morphological type.  Symbols have the same meanings as in 
Figs.~6, 7, and 13.  The dashed lines indicate the $2\sigma$ deviation 
boundaries.  
} \label{14}
\end{figure}

\begin{figure}
\vspace{60mm}
\includegraphics{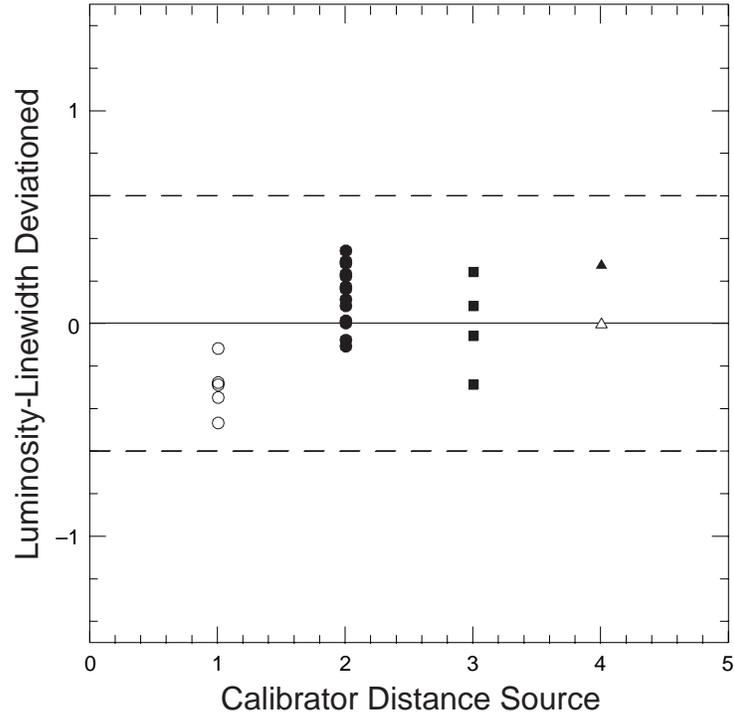}
\caption{Deviations from the $I$ band luminosity--linewidth correlation 
for the calibrator galaxies, distinguishing by source of the distance 
estimate.  Source 1 (open circles): ground based Cepheid observations;
source 2 (filled circles): HST Key Project Cepheid observations; 
source 3 (filled squares): HST Cepheid observations by first authors 
Sandage, Saha, or Tanvir reanalyzed by the Key Project Team;
source 4: HST Cepheid observations in
NGC~4603 (open triangle) and in NGC~4258 (filled triangle).
} \label{15}
\end{figure}

\end{document}